\documentclass[published]{JHEP3}

\JHEP{ }

\usepackage{cite}
\usepackage{epsf}
\usepackage{latexsym}
\usepackage{epsfig,multicol}

\def\pim{{\rm Im\,}}

\def\yzero{\smash{\hbox{$y\kern-4pt\raise1pt\hbox{${}^\circ$}$}}}
\def\p{\partial}
\def\a{\alpha}
\def\b{\beta}

\def\beq{\begin{equation}}
\def\eeq{\end{equation}}
\def\beqa{\begin{eqnarray}}
\def\eeqa{\end{eqnarray}}
\def\Om{\Omega}
\def\om{\omega}
\def\th{\theta}

\def\-{\hphantom{-}}
\def\ov{\overline}
\def\s2{\frac{1}{\sqrt2}}

\def\oh{\frac{1}{2}}
\def\beq{\begin{equation}}
\def\eeq{\end{equation}}
\def\beqa{\begin{eqnarray}}
\def\eeqa{\end{eqnarray}}

\def\IF{\relax{\rm I\kern-.18em F}}
\def\II{\relax{\rm I\kern-.18em I}}
\def\IP{\relax{\rm I\kern-.18em P}}
\def\IC{\relax\hbox{\kern.25em$\inbar\kern-.3em{\rm C}$}}
\def\IR{\relax{\rm I\kern-.18em R}}

\def\cn{{\cal N}}
\def\cam{{\cal M}}

\def\Dsl{\,\raise.15ex\hbox{/}\mkern-13.5mu D} %this one can be subscripted
\def\IZ{Z\kern-.4em  Z}

\def\inte{{\bf Z}}

\def\real{{\bf R}}
\def\T{{\bf T}}
\def\OR{\Omega {\cal R}}
\def\R{{\cal R}}
\def\ca{{\cal A}}

\def\G{\Gamma}
\def\raw{\rightarrow}

\def\Sig{\Sigma}

\def\ti{\times}

\def\IT{\bf T}

\newcommand{\drawsquare}[2]{\hbox{%
\rule{#2pt}{#1pt}\hskip-#2pt%  left vertical
\rule{#1pt}{#2pt}\hskip-#1pt%  lower horizontal
\rule[#1pt]{#1pt}{#2pt}}\rule[#1pt]{#2pt}{#2pt}\hskip-#2pt%  upper horizontal
\rule{#2pt}{#1pt}}% right vertical

\newcommand{\fund}{\raisebox{-.5pt}{\drawsquare{6.5}{0.4}}}%  fund
\newcommand{\Ysymm}{\raisebox{-.5pt}{\drawsquare{6.5}{0.4}}\hskip-0.4pt%
        \raisebox{-.5pt}{\drawsquare{6.5}{0.4}}}%  symmetric second rank
\newcommand{\Yasymm}{\raisebox{-3.5pt}{\drawsquare{6.5}{0.4}}\hskip-6.9pt%
        \raisebox{3pt}{\drawsquare{6.5}{0.4}}}%  antisymmetric second rank
\newcommand{\antifund}{\overline{\fund}}
\newcommand{\bYasymm}{\overline{\Yasymm}}

\newbox\pippobox

\title{Building MSSM Flux Vacua}

\author{Fernando Marchesano and Gary Shiu\\
      Department of Physics, 1150 University Avenue, University of Wisconsin,\\	     Madison, WI 53706, USA. \\
        E-mail: \email{marchesa@physics.wisc.edu}, \email{shiu@physics.wisc.edu}
}

\received{\today}               %%
\revised{}
\accepted{}               %% These are for published papers.

\preprint{hep-th\0409132}
\preprint{MAD-TH}

\abstract{We construct $\cn=1$ and $\cn=0$ chiral four-dimensional vacua of flux compactification in Type IIB string theory. These vacua have the common features that they are free of tadpole instabilities (both NSNS and RR) even for models with $\cn=0$ supersymmetry. In addition, the dilaton/complex structure moduli are stabilised and the supergravity background metric is warped. We present an example in which the low energy spectrum contains the MSSM spectrum with three generations of chiral matter. In the $\cn=0$ models, the background fluxes which stabilise the moduli also induce soft supersymmetry breaking terms in the gauge and chiral sectors of the theory, while satisfying the equation of motion. We also discuss some phenomenological features of these three generation MSSM flux vacua. Our techniques apply to other closed string backgrounds as well and, in fact, also allow to find new $\cn=1$ D-brane models which were believed not to exist. Finally, we discuss in detail the consistency conditions of these flux compactifications. Cancellation of K-theory charges puts additional constraints on the consistency of the models, which render some chiral D-brane models in the literature inconsistent.}


\begin{document}

\vspace*{1cm}

\section{Introduction}

Flux compactification \cite{preflux,flux1,drs,flux2,gkp} has proven to be a fruitful arena for addressing simultaneously several phenomenological issues surrounding string theory. Among the challenges in string phenomenology, moduli stabilization and supersymmetry breaking are perhaps the two most pressing ones. Standard string compactifications lead to vacua with many moduli which typically remain massless before supersymmetry is broken. An outstanding question is therefore how these moduli are lifted. Recently, there have been some attempts to stabilise moduli in string theory by considering type IIB compactifications in the presence of non-trivial RR and NSNS 3-form fluxes \cite{drs,gkp,kst}. Interestingly, depending on how the gauge and chiral sectors are embedded, these background fluxes can also induce computable soft supersymmetry breaking terms \cite{fluxsusy,lrs,ciu2}. Furthermore, compactification with fluxes provides a natural setting for warped solutions to the hierarchy problem \cite{verlinde,gkp}, along the lines of the proposal of \cite{RS}. In addition, flux compactifications have also shown to play a crucial role in the constructions of metastable de Sitter vacua \cite{kklt} and inflationary models \cite{kklmmt,inflation} from string theory. 

It is intriguing that all these phenomenologically appealing features are realised in a single framework. While this is an interesting scenario, it is important to work out some explicit models in which the full gauge and matter content of the Standard Model is embedded. With the explicit models at hand, we can then examine more quantitatively the implications of this scenario to particle phenomenology and cosmology. However, much of the discussions on flux vacua so far have focused on relatively simple models which are naturally non-chiral. The main goal of the current work is to fill in this gap and construct ${\cal N}=1$ (as well as ${\cal N}=0$ tadpole-free) supersymmetric global models of flux compactifications which admit chiral fermions. This is not as a simple task as it may seem. In fact, previous attempts seem to suggest that there is some incompatibility between chirality and tadpole cancellations (both NSNS and RR) in flux compactifications \cite{blt,cu}. However, as we shall see, not only can we obtain chiral models in flux compactifications that  are free of NSNS and RR tadpoles, but some of them are remarkably close to the Minimal Supersymmetric Standard Model (MSSM) \cite{MS}. Moreover, the non-trivial background fluxes also provide a novel way of breaking supersymmetry in a controlled manner. 

This work also has some bearing on the idea of a string landscape \cite{Susskind,Douglas}, often discussed in the context of flux vacua \cite{Douglas,taxonomy}. An implicit assumption in these statistical analysis of the string landscape is that there exist flux vacua whose low energy spectrum contains realistic features of the Standard Model. Our results provide a proof of concept that such vacua could exist. 

Recently, in \cite{MS}, we constructed the first $D=4$  Minkowski vacua of flux compactification which are (i) chiral, (ii) free of NSNS and RR tadpoles, and (iii) $\cn=1$ or $\cn=0$ supersymmetric. Moreover, we also presented in \cite{MS} the first global model of flux compactifications whose low energy spectrum contains the MSSM with three families of chiral matter.  The purpose of this paper is to elaborate the results in \cite{MS}, discuss the consistency conditions for such constructions, and describe phenomenological features of these models.

A necessary condition for the consistency of these flux vacua is the cancellation of RR tadpoles, as obtained from the familiar procedure of factorising one-loop amplitudes. However, as pointed out in \cite{angel}, this is not sufficient to ensure that the models are consistent because D-brane charges are classified by K-theory groups, rather than homology groups \cite{Ktheory}. To be precise, in addition to the homological RR charges, D-branes can also carry K-theory charges that are invisible to homology. The inconsistency arising in models with uncanceled K-theory charges is in fact more severe than that of models with uncanceled NSNS tadpoles because there is no analogue of the Fishler-Susskind mechanism for the former. In this paper, we derive the K-theory constraints on these models. These K-theory constraints are widely unnoticed in the model building literature because for simple models they are automatically satisfied. Nonetheless, these additional constraints are stringent enough to render some of the Pati-Salam-like vacua in \cite{wrong1,wrong2} inconsistent, as reflected in $SU(2)$ anomalies in these models. We hasten to stress that our models presented in \cite{MS} satisfy these K-theory constraints and hence are genuinely consistent flux vacua. 
%Contrary to the claim in \cite{wrong2}, the results in \cite{MS} were not built on the work of \cite{wrong1} and hence did not suffer from their inconsistencies.

As explained above, the main purpose of this paper is to address the construction of chiral semi-realistic flux compactifications, so let us comment on previous work in this direction. The first steps towards building chiral flux vacua were given in \cite{blt,cu}, where explicit type IIB orientifold constructions involving magnetised D-branes were studied in detail. However, when considering actual examples, some obstructions to construct $\cn=1$ models were found. Part of the task of the present paper is to explain how these difficulties can be overcomed. Nevertheless, in \cite{cu} the first example of an $\cn=0$ chiral flux compactification free of RR and NSNS tadpoles was found. This model was based on type IIB D3-branes at singularities and, although very interesting by itself, did not allow for the presence of flux-induced soft terms in the gauge sector of the theory. A more systematic approach to the embedding of the Standard Model in the flux scenario was undertaken in \cite{throat}, where special emphasis was made on the effects of warped throats and the Randall-Sundrum hierarchy that can be achieved by such embedding. These chiral constructions were again based on {\it local} models of D3-branes at singularities, following the bottom-up philosophy of \cite{botup}. Our approach is in some sense complementary to the one in \cite{throat}, since it is based on an alternative way of obtaining chirality in type IIB string theory. Namely, we focus on magnetised D-brane models, much in the spirit of \cite{blt,cu}. As we show, this approach allows us to achieve fully-fledged $D=4$, $\cn=1$ chiral flux compactifications, of which we give explicit {\it global} examples. In addition, it allows for NSNS tadpole-free $\cn=0$ chiral models, which show a more general pattern of flux-induced soft terms than that in realistic models obtained from D3-branes at singularities. Finally, there have been some attempts in building chiral compactifications with fluxes in the framework of (massive) type IIA theory \cite{massive}, leading to $\cn=1$ compactifications with $AdS_4$ minima. However, these constructions are far from realistic, both from the particle physics and the cosmological point of view \footnote{We thank Bobby Acharya and Frederik Denef for discussion on this point.}.

The paper is organised as follows. In Section 2, we describe the general strategy used to construct string theory vacua containing a chiral D-brane sector in the presence of  non-trivial RR and NSNS fluxes, and discuss the consistency conditions of such constructions. Our setup will be type IIB string theory compactified on a $\inte_2 \times \inte_2$ orientifold, containing O3 and O7-planes and magnetised D-branes \cite{blt,cu}. There are two choices of $\inte_2 \times \inte_2$ orbifolds: with or without discrete torsion. We first focus on the case with discrete torsion, which in the absence of fluxes, is T-dual to closed string background in \cite{bl,csu}. In Section 3, we describe a local MSSM model in the magnetised D-brane setup which is T-dual to the local construction in \cite{yukis}. We also discuss some phenomenological features of this construction. In Section 4, we embed this local MSSM-like model in a global flux compactification, obtaining $\cn=1$ and $\cn=0$ chiral tadpole-free models. In Section 5, we turn to the case without discrete torsion, and construct the first supersymmetric D-brane models of this closed string background, which were believed not to exist \cite{aadds}. We also show that this background admits $\cn=1$ flux vacua. We end with some final comments in Section 6. We relegate the details on the computation of the moduli space of $\cn=0$ flux vacua to Appendix A.

%\newpage

\section{Magnetised D-branes and fluxes}

In this section we review the basic results needed for building our MSSM-like flux compactifications. In particular, we describe configurations of type IIB magnetised D-branes in toroidal $\inte_2 \times \inte_2$ orientifold backgrounds, with and without the presence of 3-form fluxes. The results that we present here have been derived in \cite{blt,cu}, to where we refer the reader for more detailed discussions. In particular, both choices of a $\inte_2 \times \inte_2$ orbifold action with or without discrete torsion were discussed in \cite{blt}. In the following we will focus on constructions which involve a ${\bf T}^6/(\inte_2 \times \inte_2)$ orbifold background {\em with} discrete torsion.\footnote{The nomenclature regarding orbifolds with/without discrete torsion in the literature is somewhat confusing. In our conventions, a ${\bf T^6}/(\inte_2 \times \inte_2)$ orbifold background with discrete torsion possesses the Hodge numbers $(h_{11}, h_{21}) = (3,51)$, whereas the cohomology of the one without discrete torsion is given by  $(h_{11}, h_{21}) = (51,3)$.} This case has been thoroughly studied in \cite{cu}, whose notations and conventions we will follow. Our techniques can, nevertheless, be easily modified to the case without discrete torsion, as we show in Section \ref{extra}.

\subsection{A $\inte_2 \times \inte_2$ orientifold}

Let us first consider type IIB string theory in a closed string background given by a ${\bf T}^6 /(\inte_2 \times \inte_2)$ orbifold. Here $\T^6 = \T^2 \times \T^2 \times \T^2$ and the $\inte_2 \times \inte_2$ generators act as
\beq
\begin{array}{lcr}\vspace*{.2cm}
\th & : & (z_1, z_2, z_3) \mapsto (-z_1, -z_2, z_3)\\
\om & : & (z_1, z_2, z_3) \mapsto (z_1, -z_2, -z_3)
\end{array}
\label{genZ2}
\eeq
where $z_i$ is the complex coordinate parametrising the $i^{th}$ $\T^2$. 

Upon compactification, the massless content of the theory is given by a $D=4$, $\cn=2$ supergravity multiplet, the dilaton hypermultiplet, $h_{11}$ hypermultiplets and $h_{21}$ vector multiplets. The Hodge numbers  $(h_{11}, h_{21})$ of such background can be read as follows. On the one hand, the action  (\ref{genZ2}) projects out several components of the metric of a general $\T^6$ geometry and, as a result, we are left with fewer K\"ahler and complex structure parameters. These are encoded in the contribution of the untwisted modes of the theory to the Hodge numbers, as $(h_{11}, h_{21})_{\rm unt} = (3,3)$. On the other hand, each of the three elements $\th$, $\om$ and $\th\om$ has a fixed-point set given by 16 $\T^2$'s, and the corresponding twisted sectors do also contribute to the Hodge numbers of the orbifold. For a particular choice of discrete torsion, this contribution is given by $(h_{11}, h_{21})_{\rm tw} = (0,3 \times 16)$. 
The contributions from both the untwisted and twisted sectors
hence add up to $(h_{11}, h_{21}) = (3,51)$. 

As shown in \cite{gkp}, in order to achieve $\cn=1$ flux compactifications, it proves important to consider a type IIB background involving orientifold planes. These can be easily included in the above orbifold geometry by performing an additional modding by $\OR$, where $\Om$ is the usual world-sheet parity and $\R : z_i \mapsto - z_i$. This introduces 64 O3-planes, each one on a fixed point of $\R$, and 4 O7$_i$-planes located at the $\inte_2$ fixed points of the the $i^{th}$ $\T^2$, $i=1,2,3$, and wrapping the other two-tori. This orientifold modding projects the above $\cn=2$ spectrum to an $\cn=1$ gravity multiplet, the dilaton chiral multiplet, and 6 untwisted $+$ 48 twisted geometrical chiral multiplets.

\subsection{with fluxes}

The above orientifold geometry can be generalised to a more involved type IIB closed string background. Indeed, to the above compactification we could think of adding a discrete internal B-field \cite{bflux} or NSNS and RR field strength fluxes. In this paper we will be mainly interested in the second possibility. Indeed, type IIB compactifications on Calabi-Yau threefolds $\cam_6$ with non-trivial 3-from field strength backgrounds have been extensively studied in the literature \cite{flux1,drs,flux2,gkp,kst,fp,flux3,gauged,kt,blt,cu,afterflux}, providing interesting examples of moduli stabilization \cite{drs,gkp,kst,kt} and soft supersymmetry breaking \cite{fluxsusy,lrs,ciu2}. One of the main goals of this paper is to provide examples where both phenomena occur at the same time and in a controlled manner.

Given type IIB string theory compactified on a Calabi-Yau threefold $\cam_6$, it is possible to introduce both RR and NSNS 3-form field strength backgrounds $F_3$ and $H_3$. The results in \cite{gkp} provide, in a quite general setup, the conditions that the fluxes must satisfy in order to yield consistent $D=4$ Minkowski compactifications. In particular, they must obey the Bianchi identities
\beq
dF_3 = 0 \quad \quad dH_3 = 0
\label{bianchi}
\eeq
and must be properly quantised over any 3-cycle $\Sig \subset \cam_6$, hence defining an integer cohomology class in $H^3(\cam_6,\inte)$. These conditions simplify considerably if one considers $\cam_6$ to be a toroidal orbifold ${\bf T}^6/\G$, and the 3-form fluxes to belong to the untwisted cohomology\footnote{One may in principle turn on fluxes on the orbifold twisted 3-cycles. See, e.g. the example in \cite{blt}. The supergravity approximation where most of the flux compactification physics is based on, however, is no longer reliable in this case.}. Indeed, in this case we can simply take $F_3$ and $H_3$ to be constant 3-forms, while the quantisation conditions amount to
\beq
{1 \over (2 \pi)^2 \a^\prime} \int_{\Sigma} F_3 \in N_{\rm min} \cdot \inte, \quad \quad
{1 \over (2 \pi)^2 \a^\prime} \int_{\Sigma} H_3 \in N_{\rm min} \cdot \inte
\label{quant}
\eeq
where $\Sig$ is a 3-cycle on the covering space ${\bf T}^6$, and $N_{\rm min}$ is an integer number which depends on the orbifold group $\G$.\footnote{Roughly speaking, $N_{\rm min}$ can be computed from dividing the volume of a 3-cycle on $\cam_6$ by its corresponding fractional cycle on $\cam_6/\G$. See \cite{blt} for a more detailed discussion on this point.} In the particular $\inte_2 \times \inte_2$ orbifold with discrete torsion we are discussing $N_{\rm min} = 4$. Considering an $\OR$ orientifold of this background implies that we have an additional $\inte_2$ subgroup acting on ${\bf T}^6$, and hence the previous number $N_{\rm min}$ is effectively\footnote{The whole truth is a little more subtle \cite{fp}. We may either have that the flux quanta along the cycle $\Sig$ are of the form $(2n+1)N_{\rm min}$ or $2n N_{\rm min}$, with $n \in \inte$, depending on the fact that $\Sig$ passes over an odd or even number of O3$^{(+,+)}$. In the $\inte_2 \times \inte_2$ orientifold at hand there are no such exotic O3-planes, and so we can take $N_{\rm min}^{\rm ori} = 2 N_{\rm min}^{\rm orbi}$. \label{subtle}} multiplied by $2$. Hence, at the end of the day, we must impose that the fluxes $F_3$, $H_3$ are quantised in multiples of 8 over toroidal cycles.

In type IIB supergravity language, these 3-form fluxes are more elegantly encoded in the complexified 3-form flux
\beq
G_3 = F_3 - \tau H_3
\label{G3}
\eeq
where $\tau = a + i/g_s$ is the usual axion-dilaton coupling. By looking at the supergravity action it is easy to see that these fluxes both gravitate and couple to the $C_4$ RR potential, hence carrying both D3-brane charge and tension. In D3-brane units, the RR charge is given by
\beq
N_{\rm flux} = {1 \over (4\pi^2 \a^\prime)^2} \int_{\cam_6} H_3 \wedge F_3 = {i \over (4\pi^2 \a^\prime)^2} \int_{\cam_6} {G_3 \wedge \overline{G}_3 \over 2\pim \tau}
\label{RRcharge}
\eeq
which is a topological quantity that, if we impose conditions (\ref{quant}), is an integer number proportional to $(N_{\rm min})^2$. The D3-brane tension is, on the other hand, given by
\beq
T_{\rm flux}  =  - {1 \over (4\pi^2 \a^\prime)^2} \int_{\cam_6} {G_3 \wedge *_6 \overline{G}_3 \over 2\pim \tau} \geq |N_{\rm flux}|
\label{tensionflux}
\eeq
which is not topological but depends on the geometry of the compactification. More precisely, it depends on the dilaton and complex structure moduli of $\cam_6$ and, as a matter of fact, the quantity $V_{\rm eff} = T_{\rm flux} - |N_{\rm flux}|$ can be thought of an effective potential for those moduli \cite{kst}. The minimum of such potential is reached when the BPS-like bound in (\ref{tensionflux}) is saturated, which amounts to imposing the ISD ($*_6 G_3 =  i G_3$) or IASD ($*_6 G_3 = - i G_3$) conditions. Attaining such conditions drives the dilaton/complex structure moduli to a particular value, `lifting' those compactification moduli \cite{drs,gkp}. From (\ref{RRcharge}) and (\ref{tensionflux}) is easy to see that an ISD flux carries the charge and tension of $|N_{\rm flux}|$ D3-branes, whereas an IASD flux would carry those of $|N_{\rm flux}|$ anti-D3-branes. In the $\inte_2 \times \inte_2$ orientifold we are discussing, however, introducing $\overline{D3}$-branes/IASD fluxes would lead to models with broken supersymmetry and uncanceled NSNS tadpoles, which make it difficult to solve the supergravity equations of motion, create run-away potentials for some moduli and typically break $D=4$ Poincar\'e invariance \cite{NSNS}. We will thus restrict to models containing ISD 3-form fluxes, much in the spirit of \cite{gkp}.

An interesting point of flux compactifications is that, although any ISD flux carries the tension and charge of a D3-brane, not any ISD flux will preserve the same amount of supersymmetry that a D3-brane would do. Indeed, in a generic ${\bf CY_3}$ compactification, where the holonomy of $\cam_6$ is contained in SU(3) but not a proper subgroup of it, there is a unique covariantly constant spinor and hence a preferred complex structure. We can decompose the 3-form flux $G_3$ in terms of the Hodge cohomology according to such complex structure, finding that if $G_3$ is an ISD flux it must consist of (2,1) and (0,3)-forms. By the results of \cite{granapol}, one finds that only those ISD fluxes which consist of (2,1) components do preserve $\cn=1$ supersymmetry\footnote{The precise result states that $G_3$ must be a (2,1) {\it primitive} flux in order to preserve $\cn=1$. The condition of primitivity is nevertheless automatically satisfied by any 3-form in a generic ${\bf CY_3}$ compactification. For generalisations of this result see \cite{ansatze}.}. Indeed, an ISD $G_3$ flux with a non-vanishing (0,3) component would break an $\cn=1$ compactification down to $\cn=0$, giving a mass to the gravitino of the form
\beq
m_{3/2}^2 \sim {\left|\int G_3 \wedge \Om \right|^2 \over \pim \tau \ {\rm Vol }(\cam_6)^2}.
\label{gravitino}
\eeq
The cosmological constant of the compactification, however, remains zero 
(to lowest order) due to the no-scale structure of the scalar potential $V_{\rm eff}$ \cite{gkp}. 

It is important to notice that, due to the backreaction of the flux and other objects such as D-branes and O-planes, the $\cam_6$ compactification metric is not actually Calabi-Yau, but rather conformally equivalent to it. The difference comes from a non-trivial warp factor, which only depends on the internal components of $\cam_6$ and hence preserves $D=4$ Poincar\'e invariance \cite{drs,gkp}. In the above discussion we have neglected this warping, although the qualitative features we have described remain valid when we include its effects. On the other hand, such warped geometries have been considered as a natural way to generate hierarchies in string theory constructions, realising the Randall-Sundrum scenario \cite{gkp}, as well as a necessary ingredient in the recent proposal for constructing de Sitter string vacua \cite{kklt}.

\subsection{and magnetised D-branes}

As usual, a closed string background with orientifold planes generates a non-trivial contribution to the Klein bottle string amplitude, and hence crosscap tadpoles. These can be cancelled by introducing an open string sector in the theory. In the case at hand, this open string sector will consist of type IIB D$(3+2n)$-branes, filling up $D=4$ Minkowski space and wrapping $2n$-cycles on the compact manifold. In the following we will describe the consistency conditions that such D-brane models must satisfy in order to yield $\cn = 1$ chiral compactifications, as well as the properties of the latter.

First, let us point out that, if we forget about fluxes, many $\cn=1$ models of the above kind have already been built in the literature. Indeed, when no 3-form fluxes are turned on, this $\inte_2 \times \inte_2$ orientifold background is related by T-duality to the $\cn = 1$ constructions in \cite{bl,csu}. More precisely, the T-dual of the Type I vacua in \cite{bl} would be achieved by adding 32 D3-branes and 32 D7$_i$-branes of each kind, whereas the analogue of the intersecting D-brane constructions in \cite{csu} corresponds to introducing D9, D7 and D3-branes with magnetic fluxes. This last construction is particularly interesting, since it allows for $D=4$, $\cn=1$ chiral compactifications, chirality arising from open strings \cite{bachas,bdl}. 

Let us describe in some detail these magnetised D-brane models. The topological information of a set $a$ of D9-branes is encoded in seven integer numbers. The number of D9-branes $N_a$, and the six `magnetic numbers' $(n_a^i,m_a^i)$, $i=1,2,3$. Here $m_a^i$ is the number of times the D9's are wrapped on the $i^{th}$ ${\bf T}^2$, and $n_a^i$ the unit of magnetic flux on such two-torus\footnote{For each fixed value of $a$ and $i$, $n_a^i$, $m_a^i$ are assumed to be coprime numbers.}. More precisely, we are turning on a constant Abelian world-volume magnetic field $F=dA$, satisfying
\beq
{m_a^i \over 2\pi} \int_{{\bf T}^2_i} F_a^i = n_a^i.
\label{magnet}
\eeq
This notation also allows us to describe D-branes of lower dimension \cite{cu}, so that we have
\beq
\begin{array}{lclccc}\vspace*{.2cm}
& & & ({\bf T}^2)_1 & ({\bf T}^2)_2 & ({\bf T}^2)_3
\\ \vspace*{.2cm}
D9 & \raw & N_a \quad & (n_a^1,m_a^1) & (n_a^2,m_a^2) & (n_a^3,m_a^3)
\\\vspace*{.2cm}
D7_1 & \raw & N_a \quad & (1,0) & (n_a^2,m_a^2) & (n_a^3,m_a^3)
\\\vspace*{.2cm}
D5_1 & \raw & N_a \quad & (n_a^1,m_a^1) & (1,0) & (1,0)\\
D3 & \raw & N_a \quad &(1,0) & (1,0) & (1,0)
\end{array}
\label{table}
\eeq
where D5$_i$ stands for a D5-brane wrapping the $i^{th}$ two-torus $({\bf T}^2)_i$ and D7$_i$ a D7-brane {\em not} wrapping $({\bf T}^2)_i$. This description also shows that the presence of a worldvolume magnetic flux on a D-brane induces D-brane charge of lower dimension \cite{induced}. For instance, a D9-brane with magnetic flux will usually have also charges of D7, D5 and D3-brane. This fact is quite relevant when satisfying RR tadpoles conditions \cite{aads,afiru}, and will prove essential for finding $\cn=1$ flux compactifications in this setup. In general, this topological information is encoded in a charge class $[{\bf Q}_a]$, which describes the full set of RR charges of the D-brane sector $a$.

Now, since we are in an orientifold theory, the whole configuration must be invariant under the $\inte_2 \times \inte_2 \times \OR$ orientifold group. This means that the stack $a$ of $N_a$ D-branes is either invariant under such geometrical action or is accompanied by its orientifold images. In particular, the orientifold group has a natural action on the magnetic numbers $(n_a^i,m_a^i)$, given by
\beq
\begin{array}{ccl}
\inte_2 \times \inte_2 & : & (n_a^i,m_a^i)\ \mapsto\ (n_a^i,m_a^i)\\
\OR & : & (n_a^i,m_a^i)\ \mapsto\ (n_a^i,-m_a^i)\\
\end{array}
\label{topaction}
\eeq
Notice that, even if the $\inte_2 \times \inte_2$ orbifold group acts trivially on the topological magnetic numbers, this does not necessarily means that the stack $a$ is invariant under it. This will depend on the moduli of the configuration, namely the positions and Wilson lines of each D-brane sector. The $\OR$ image of the D-brane stack $a$, which usually has different magnetic numbers, is denoted by $a'$.

Finally, it is useful to describe the orientifold plane content of the theory in terms of magnetic numbers. Indeed, we can express the set of O3-planes and D7$_i$-planes in terms of their RR charges, properly normalised in D-brane units. From this we recover that the total orientifold charge is given by $-32$ times the charge class 
\beqa \nonumber
[{\bf Q}_O] & = & [(1,0)\ (1,0)\ (1,0)] \ +\  [(1,0)\ (0,1)\ (0,-1)] \\
& + & [(0,1)\ (1,0)\ (0,-1)] \ +\  [(0,1)\ (0,-1)\ (1,0)].
\label{oriclass}
\eeqa

The topological magnetic numbers also allows us to compute the chiral spectrum of the theory in a simple way. Let us momentarily forget about the orientifolded geometry. A stack of $N_a$ D-branes on ${\bf T}^6$ would give rise, after dimensional reduction to $D=4$, a $U(N_a)$ gauge group from the $aa$ sector, i.e., from open strings with both ends on the same stack $a$. A pair of stacks $a$ and $b$ would then yield the gauge group $U(N_a) \times U(N_b)$, as well as $I_{ab}$ chiral fermions charged in the bifundamental representation $(N_a, \bar{N}_b)$, where
\beq
I_{ab} = [{\bf Q}_a] \cdot [{\bf Q}_b]= \prod_{i=1}^{3} \left(n_a^i m_b^i - m_a^i n_b^i \right)
\label{intersection}
\eeq
is the intersection product of charge classes\footnote{Recall that in general $I_{ab} \in \inte$, and that positive and negative intersection products correspond to fermions of opposite chirality.}. An explicit field theory derivation of this fact has been given in \cite{magnus}. 

This general picture also works in the orientifold case, which nevertheless adds some new features. Just as in \cite{csu}, we will mainly build our models from D-branes which are fixed by some element of the $\inte_2 \times \inte_2$ orbifold group\footnote{This fact is indeed necessary in order to get a odd number of generations.}. If the magnetic numbers are not fixed by the $\OR$ action, we will still have a unitary gauge group. The orbifold action will nevertheless project the initial $U(N_a)$ Chan-Paton gauge group down to $U(N_a/2)$. In addition to this $\cn=1$ vector multiplet, the massless $aa$ sector will consist of three chiral multiplets transforming in the adjoint representation, which can be understood as the open string moduli (positions and/or Wilson lines) of the stack $a$ of D-branes. This and the chiral spectrum arising from other sectors
are summarised in table \ref{matter}.
\TABLE{\renewcommand{\arraystretch}{1.25}
\begin{tabular}{|c|c|}
\hline
\hspace{1cm} {\bf Sector} \hspace{1cm} &
\hspace{1cm} {\bf Representation} \hspace{1cm} \\
\hline\hline
$aa$   &  $U(N_a/2)$ vector multiplet  \\
       & 3 Adj. chiral multiplets   \\
\hline\hline
$ab+ba$   & $I_{ab}$ $(\fund_a,\antifund_b)$ chiral multiplets  \\
\hline\hline
$ab'+b'a$ & $I_{ab'}$ $(\fund_a,\fund_b)$ chiral multiplets  \\
\hline\hline
$aa'+a'a$ & $\frac 12 (I_{aa'} - 4 I_{a,O}) \;\;
\Ysymm\;\;$ chiral  multiplets \\
          & $\frac 12 (I_{aa'} +  4 I_{a,O}) \;\;
\Yasymm\;\;$ chiral multiplets \\
\hline
\end{tabular}
\label{matter}
\caption{\small Massless spectrum for $\cn=1$ magnetised D-branes in the $\IT^6/(\inte_2\times \inte_2)$ $\OR$ orientifold. $I_{a,O}$ stands for the intersection product between $[{\bf Q}_a]$ and the orientifold plane charge class $[{\bf Q}_O]$.}}

When $[{\bf Q}_a]$ is invariant under the $\OR$ action, which is the case for D3 and D7$_i$-branes without magnetic fluxes, the gauge group will no longer be unitary but symplectic. In particular, if we take a stack of $2N_a$ D-branes with $\OR [{\bf Q}_a] = [{\bf Q}_a]$ we will recover a $USp(N_a)$ gauge group. As in the case of unitary gauge groups, this implies that $N_a$ has to be an even number. Besides, whenever $N_a$ is a multiple of 4, we can Higgs this gauge group as $USp(4n) \raw U(1)^n$ \cite{bl,csu}. The corresponding spectrum is presented in Table \ref{matter2}.
\TABLE{\renewcommand{\arraystretch}{1.25}
\begin{tabular}{|c|c|}
\hline
\hspace{1cm} {\bf Sector} \hspace{1cm} &
\hspace{1cm} {\bf Representation} \hspace{1cm} \\
\hline\hline
$aa$   &  $USp(N_a)$ vector multiplet  \\
       & 3 $\Yasymm\;\;$ chiral multiplets   \\
\hline\hline
$ab+ba$   & $I_{ab}$ $(\fund_a,\antifund_b)$ chiral multiplets  \\
\hline
\end{tabular}
\label{matter2}
\caption{\small Massless spectrum for D-branes invariant under the orientifold action (i.e., $\OR [{\bf Q}_a] = [{\bf Q}_a]$). Notice that now $ab$ and $b'a$ sectors are identified.}}

Given this gauge group and chiral spectrum, one may consider building semi-realistic chiral models close to the Standard Model or any extension of it. There are, however, a set of consistency conditions which any string compactification must satisfy, known as tadpole cancellation conditions. First, cancellation of RR tadpoles implies that the homological RR charges of the D-branes and the O-planes cancel, that is
\beq
\sum_\a N_\a \left( [{\bf Q}_\a] + [{\bf Q}_{\a'}] \right) = 32 [{\bf Q}_O]
\label{RR}
\eeq
which can be reexpressed as
\beq
\begin{array}{rcl} \vspace*{.2cm}
\sum_\a N_\a n_\a^1 n_\a^2 n_\a^3 & = & 16, \\\vspace*{.2cm}
\sum_\a N_\a m_\a^1 m_\a^2 n_\a^3 & = & -16, \\\vspace*{.2cm}
\sum_\a N_\a m_\a^1 n_\a^2 m_\a^3 & = & -16, \\\vspace*{.2cm}
\sum_\a N_\a n_\a^1 m_\a^2 m_\a^3 & = & -16.
\end{array}
\label{tadpoles}
\eeq
where we are not summing over orientifold images. These conditions are directly related to cancellation of $D=4$ chiral anomalies arising in triangular diagrams, such as $SU(N)^3$ and $U(1)$ anomalies. 

There is, however, a subtle issue regarding RR tadpole cancellation, which was pointed out in \cite{cu} but is usually ignored in the model-building literature. In presence of orientifold planes, D-branes may carry discrete charges which are invisible from the point of view of one-loop divergences and the supergravity Bianchi identities, but which are classified by $\inte_2$ K-theory groups \cite{Ktheory}. Now, in order to build a consistent string theory model any RR D-brane charge in our compact space must vanish, including these exotic K-theory charges. Cancellation of discrete K-theory charges will in general impose extra consistency constraints, in addition to the familiar RR tadpole conditions based on homological charges and which in our case are given by (\ref{tadpoles}). Notice that these new K-theory constraints do not have any effect on, e.g., $D=4$ $SU(N)^3$ chiral anomalies, which already vanish by imposing (\ref{tadpoles}). Following \cite{angel}, however, we can see their effect by introducing suitable probes on the theory. Indeed, let us consider D3 and D7$_i$-branes without magnetic fluxes, giving rise to $USp(2) \simeq SU(2)$ gauge groups. Although free of cubic anomalies, an $SU(2)$ group suffers from a global gauge anomaly if there is an odd number of $D=4$ fermions charged in the fundamental representation \cite{SU(2)}. In the present context, requiring the absence of such global anomalies amounts to imposing the constraints\footnote{These constraints are only valid for $\inte_2 \times \inte_2$ orientifolds without discrete B-field. In the presence of a non-vanishing B-field they usually weaken. See \cite{tesis}, Appendix B, for a detailed study of these K-theory constraints, with and without B-field, in the case of toroidal orientifolds.}
\beq
\begin{array}{rcl} \vspace*{.2cm}
\sum_\a N_\a m_\a^1 m_\a^2 m_\a^3 & \in & 4 \inte, \\\vspace*{.2cm}
\sum_\a N_\a n_\a^1 n_\a^2 m_\a^3 & \in & 4 \inte, \\\vspace*{.2cm}
\sum_\a N_\a n_\a^1 m_\a^2 n_\a^3 & \in & 4 \inte, \\\vspace*{.2cm}
\sum_\a N_\a m_\a^1 n_\a^2 n_\a^3 & \in & 4 \inte.
\end{array}
\label{tadpolesK}
\eeq
Notice that these are indeed $\inte_2$ charge constraints, since $N_\a$ are already even integers. In K-theory language, we are imposing the global cancellation of $\inte_2$ RR charges, carried by fractional $D5_i -\overline{D5}_i$ and $D9-\overline{D9}$ pairs.

Second, we would like to construct models free of NSNS tadpoles, that is, such that the tensions of the objects in the configuration do also cancel. In a magnetised D-brane configuration with vanishing RR tadpoles, this can be achieved by requiring that every set of D-branes preserves the same $\cn=1$ supersymmetry unbroken by the orientifold. This usually implies a condition on the K\"ahler parameters, which in the present context reads\footnote{This formula is actually only valid for the case $n_a^i \geq 0$. See below for some other important cases.}
\beq
\sum_i {\rm tan}^{-1} \left( {m_a^i \ca_i \over n_a^i} \right) = 0,
\label{susy}
\eeq
where $\ca_i$ is the area of $({\bf T}^2)_i$ in $\a'$ units. A small deviation from this condition can be understood as a non-vanishing FI-term in the $D=4$ effective theory \cite{csu,cim1}.

\subsection{yielding $\cn=1$ and $\cn=0$ chiral compactifications} 

In the presence of RR and NSNS fluxes, magnetised D-brane configurations may yield new and interesting phenomena. First, as usual from flux compactification, the scalar potential fixes most of the dilaton/complex structure moduli. Second, since the flux carries charge and tension of D3-brane, the RR and NSNS tadpole conditions get modified. In the $\inte_2 \times \inte_2$ orientifold discussed above, the most general form of an (untwisted) ISD flux is given by\footnote{For the sake of clarity, we are dropping the $(4\pi^2\a')^{-1}$ normalisation factor in the definition of the flux.}
\beqa
G_3 & = & G_{\bar{1}23} \, d{\ov z}_1dz_2dz_3 + G_{1\bar{2}3} \, dz_1d{\ov z}_2dz_3 + G_{12\bar{3}} \, dz_1dz_2d{\ov z}_3 + G_{\bar{1}\bar{2}\bar{3}} \,d{\ov z}_1 d{\ov z}_2 d{\ov z}_3.
\label{ISD}
\eeqa
The D3-brane RR charge carried by this flux is given by (\ref{RRcharge}) which can be written as
\beq
N_{\rm flux} = 4 g_s \pim \tau_1  \pim \tau_2 \pim \tau_3 \,
\left(|G_{\bar{1}23}|^2 + |G_{1\bar{2}3}|^2 + |G_{12\bar{3}}|^2 + |G_{\bar{1}\bar{2}\bar{3}}|^2 \right)
\label{RRcharge2}
\eeq
where $\tau_i$ is the complex structure modulus of $(\T^2)_i$, usually fixed by the ISD condition. This modifies the RR tadpole conditions (\ref{tadpoles}), whose first line now reads
\beq
\sum_\a N_\a n_\a^1 n_\a^2 n_\a^3 + \oh N_{\rm flux} = 16,
\label{tadpoleflux}
\eeq
while the rest of the tadpole conditions in (\ref{tadpoles}) and (\ref{tadpolesK}) remain unchanged.

Thirdly, ISD fluxes may lead to $\cn=0$ models with softly broken supersymmetry and free of NSNS tadpoles. Indeed, it has been recently realised \cite{lrs,ciu2} that a non-vanishing $(0,3)$ component of $G_3$ not only does give mass to the gravitino, but also induces soft terms in the worldvolume of D7-branes implying, e.g., a non-vanishing mass for the gauginos and the scalar fields in the adjoint. The same effect occurs in D3-D7 and D7-D7 chiral sectors, where the scalar components of the chiral multiplets would also get a mass. These effects imply that ISD fluxes not only stabilise closed string moduli, but can also give masses to open string moduli, in agreement with the results in \cite{gktt,cali}. In addition, we learn that, in the presence of a non-vanishing $G_{\bar{1}\bar{2}\bar{3}}$ flux component, the $\cn=1$ spectrum of table \ref{matter} is softly broken to an $\cn=0$ chiral spectrum. Moreover, the soft term structure derived from flux compactification turns out to be particularly simple, which has suggested interesting solutions to several problems associated with soft SUSY breaking patterns \cite{fluxed}.

Finally, let us also mention that many other subtle issues, mainly associated with D-branes in the presence of a NSNS flux $H_3$ do also appear in this class of compactifications. For instance, the presence of 3-form fluxes modifies the usual anomaly cancellation pattern \cite{angel2}, and brane-bulk mixed anomalies could in principle appear \cite{CP}. In addition, the presence of a non-vanishing $H_3$ modifies the K-theory group of the theory \cite{Ktheory}. This may lead, e.g., to instanton processes violating homological D-brane charge \cite{mms}. These issues were analysed in \cite{cu} with the outcome that, at least in this particular class of constructions, they do not affect either the chiral spectrum of tables \ref{matter} and \ref{matter2} or the consistency conditions (\ref{tadpoles}) and (\ref{tadpolesK}), except for (\ref{tadpoleflux}).

\section{A local MSSM-like model}

In the last few years, it has been shown that string theory constructions based on intersecting D-branes could lead to new chiral $D=4$ compactifications with semi-realistic spectra and many phenomenologically appealing features \cite{bgkl,afiru,afiru2,bkl,imr}. In particular, in \cite{more,yukis} a local intersecting D6-brane model, with the gauge group and $\cn=1$ chiral spectrum of the MSSM, was presented. The construction was based on a simple $\T^6$ orientifold background, which, due to its limitations, forbids any $\cn=1$ chiral construction \cite{bgkl}.\footnote{As pointed out in \cite{yukis}, the intersecting D-brane structure of this local model in principle allows for an $\cn=1$ embedding on a more general background, such as a ${\bf CY_3}$. This was indeed shown in \cite{dhs}, where $\cn=1$ vacua based on Gepner model orientifolds and these intersection numbers were found.} One of the purposes of this paper is to embed this local MSSM-like construction in a fully-fledged $\cn=1$ string compactification, which moreover shares the nice and simple properties of the toroidal construction.
\footnote{There have been previous attempts in this direction. Unhappily, those models do not satisfy RR tadpole conditions (\ref{tadpolesK}), and so cannot be thought of as consistent string theory constructions.}
 As we will see, the  $\inte_2 \times \inte_2$ magnetised D-brane setup allows for such an embedding and, moreover, does also admit the presence of ISD fluxes.% in the compactification.

\TABLE{\renewcommand{\arraystretch}{1.25}
\begin{tabular}{|c||c|c|c|}
\hline
 $N_\a$  &  $(n_\a^{1},m_\a^{1})$  &  $(n_\a^{2},m_\a^{2})$
&  $(n_\a^{3},m_\a^{3})$ \\
\hline\hline $N_a = 6$ & $(1,0)$ & $(g,1)$ & $(g,-1)$  \\
\hline $N_b=2$ & $(0,1)$ &  $ (1,0)$  & $(0,-1)$ \\
\hline $N_c=2$ & $(0,1)$ &  $(0,-1)$  & $(1,0)$  \\
\hline $N_d = 2$ & $(1,0)$ & $(g,1)$ & $(g,-1)$  \\
\hline \end{tabular}
\label{MSSMblock}
\caption{\small D-brane magnetic numbers giving rise to an $SU(3) \times SU(2) \times SU(2) \times U(1)$ extension of the MSSM with $g$ generations of chiral multiplets.}} 
Let us first review the features of this local construction. In absence of 3-form fluxes, the D-brane content of the MSSM-like model in \cite{yukis} can be translated to magnetised D-brane model by simple T-duality, and then embedded into a $\inte_2 \times \inte_2$ geometry. The result is shown in table \ref{MSSMblock}.
The D-brane content of this model consist of four stacks of D7-branes. More precisely, we consider two stacks of D7$_2$ and D7$_3$-branes without magnetic fluxes, giving rise to an $SU(2) \times SU(2)$ gauge group via the spectrum of table \ref{matter2} and the identification $USp(2) \simeq SU(2)$. In addition, we have two stacks of D7$_1$-branes with non-trivial magnetic fluxes, yielding a $U(3) \times U(1)$ gauge group. The chiral spectrum of this model can be easily computed from the intersection products of the D7-branes, given by $I_{ab} = I_{db} = g$ and $I_{ac} = I_{dc} = -g$. We thus recover a Left-Right extension of the MSSM gauge group, and with $g$ generations of chiral multiplets. This spectrum is displayed in table \ref{LRMSSM}, together with their identification with matter fields, for the particular value $g=3$.

\TABLE{\renewcommand{\arraystretch}{1.2}
\begin{tabular}{|c|c|c|c|c|c|}
\hline Sector &
 Matter fields  & $SU(3) \ti SU(2)_L \ti SU(2)_R $  &  $Q_a$  & $Q_d$  & $Q_{B-L}$ \\
\hline\hline (ab) & $Q_L$ &  $3(3,2,1)$ & 1   & 0 & 1/3 \\
\hline (ac) & $Q_R$   &  $3( {\bar 3},1,2)$ & -1  &  0  & -1/3 \\
\hline (db) & $L_L$    &  $3(1,2,1)$ &  0 & 1  & -1  \\
\hline (dc) & $L_R$   &  $3(1,1,2)$ &  0  & -1  &  1  \\
\hline 
\end{tabular}
\label{LRMSSM}
\caption{\small Left-Right MSSM spectrum and $U(1)$ charges obtained from table \ref{MSSMblock}, for the particular choice $g=3$. The B-L generator is defined as $Q_{B-L} = \frac 13 Q_a - Q_d$.}}
Let us point out several interesting features of this construction:

\begin{itemize}

\item
It is easy to check that this spectrum is free of cubic chiral anomalies, whereas it develops an anomalous $U(1)$ given by $U(1)_a + U(1)_b$. Such anomaly is cancelled by means of a generalised Green-Schwarz mechanism \cite{afiru}, which in turn gives a Stueckelberg mass to the $U(1)$ gauge boson. We are thus left with a gauge symmetry group given by just $SU(3) \times SU(2)_L \times SU(2)_R \times U(1)_{B-L}$, and where the Baryon and Lepton numbers remain as perturbative global symmetries of the effective theory. 

\item
Notice, as well, that the stacks $a$ and $d$ posses the same magnetic numbers $[{\bf Q}_a] = [{\bf Q}_d]$. This allows to put them on top of each other, where the enhancement $SU(3) \times U(1)_{B-L} \raw SU(4)$ occurs. At some particular points of the open string moduli space we then recover a full Pati-Salam spectrum. In fact, from particle physics perspective, this D-brane local model can be thought of as a Pati-Salam broken to a Left-Right symmetric MSSM by means of $SU(4)$ adjoint Higgsing \cite{yukis}. The v.e.v.'s of the three adjoint fields are given, in the stringy geometrical language, by the relative position of the stacks $a$ and $d$ on $(\T^2)_1$, and the relative Wilson lines on $(\T^2)_2$, $(\T^2)_3$.

\item
Another characteristic feature of this local model is the Higgs sector, which arises between strings stretched between stacks $b$ and $c$. Notice that the intersection product $I_{bc}$ vanishes, which does not mean that there are no charged particles in the $bc$ sector, but only that the net chirality is zero. Indeed, this is a well-known feature of the MSSM Higgs sector, and in particular of its Left-Right extension. It is easy to adapt the computations in \cite{bl} to this T-dual picture, and check that $bc$ strings provide one $\cn=1$ chiral multiplet in the $(2,2)$ representation of $SU(2)_L \times SU(2)_R$. We hence recover the Higgs sector of (the LR extension of) the MSSM, which moreover comes equipped with a $\mu$-term. Indeed, the open string tree-level superpotential studied in \cite{bl} give us the $SU(2)_L \times SU(2)_R$ invariant coupling
\beq
W = \left( \langle \zeta_{cc}^1 \rangle - \langle \zeta_{bb}^1 \rangle \right) \cdot {\rm det\ } {\bf H}
\label{muterm}
\eeq
where $\zeta_{bb}^1$, $\zeta_{cc}^1$ are the complex singlets arising from $(\T^2)_1$ and from sectors $bb$ and $cc$, respectively, and ${\bf H}$ is the $2 \times 2$ matrix of the Higgs fields\footnote{In \cite{bl} this system is given by a D5$_i$D5$_j$ sector, and $\zeta_{bb}^1$, $\zeta_{cc}^1$ have the interpretation of D5-brane positions.}.

\item
Finally, the Yukawa couplings between the chiral fields of this model are extremely simple and can be elegantly expressed in terms of theta functions. This was explicitly computed in \cite{yukis} and more recently in \cite{magnus}, where techniques for computing Yukawa couplings in magnetised D-brane models were developed.
%\footnote{The computations in these references are developed in the framework of toroidal orientifolds, but is easy to see that the final expressions are also valid for other orientifolds, as the $\inte_2 \times \inte_2$ considered here.}
 Indeed, following these references one can diagonalise analytically the chiral matter mass matrix, obtaining one massive and two massless generations of quarks and leptons.

\end{itemize}

Of course, in order to provide a consistent $\cn=1$ string construction, we need to satisfy both RR and NSNS tadpole cancellation conditions. It is easy to see that the $\cn=1$ conditions (\ref{susy}) are satisfied by imposing
\beq
\ca_2 = \ca_3
\label{susyMSSM}
\eeq
upon which we recover an $\cn=1$ spectrum in the chiral sector of the theory. On the other hand, while conditions (\ref{tadpolesK}) are satisfied by table \ref{MSSMblock}, this is not the case for RR conditions (\ref{tadpoles}), for any value of $g$. This implies that we need to add extra RR and NSNS sources, such as more D-branes and 3-form fluxes, in order to construct a fully-fledged tadpole-free string compactification. We will perform such completion in the next section, where an explicit example will be constructed. Meanwhile, we can think of the D-brane content of table \ref{MSSMblock} as a MSSM-like block, and study its local properties.

\subsection{From Left-Right to MSSM}

One obvious question to ask if this $\cn=1$ Left-Right symmetric model can actually be broken to a MSSM spectrum. Unlike its toroidal version \cite{yukis}, in the $\inte_2 \times \inte_2$ construction $SU(2)_R$ cannot be broken down to $U(1)$ by adjoint Higgsing. There is however another possibility, more difficult to visualise geometrically but equally valid from the field theory viewpoint. This basically amounts to applying the general ideas in \cite{cim2} to this particular example, identifying the non-adjoint, rank reduction, Higgsing with the process of D-brane recombination.

Let us see how this works in some detail. Recall that in the sector $dc$ we have three chiral multiplets with the quantum numbers $(1,1,2)_{-1}$, which are the perfect candidates for breaking $SU(2)_R \times U(1)_{B-L} \raw U(1)_Y$. This intuition can be checked in the D-brane picture. Indeed, giving a v.e.v. to the massless scalars in these multiplets amounts to taking a flat direction in the superpotential, which corresponds to the recombination process
\beq
\begin{array}{rcl}
c+d & \raw & j \\
\left. 2 [{\bf Q}_c] + 2 [{\bf Q}_d] \right. & \raw &  2 [{\bf Q}_j] = 2 [{\bf Q}_c + {\bf Q}_d]
\end{array}
\label{recomb1}
\eeq
Notice that, after this recombination, $[{\bf Q}_j]$ cannot be simply written in terms of the magnetic numbers $\{(n^i,m^i)\}_{i=1}^3$, i.e., is no longer a `factorisable' D-brane but a BPS bound state of two of them \cite{bound}. Notice as well that  $[{\bf Q}_j] \neq  \OR [{\bf Q}_j]$, so that we recover a $U(1)_j$ gauge group after this recombination.

Although hard to visualise, we can still compute the intersection product between the stack $j$ and the other D-branes in the local model. We obtain
\beq
\begin{array}{lcl}
I_{ab} = g \\
I_{aj} = -g & & I_{aj'} = -g \\
I_{bj} = -g \\
I_{jj'} = -2g
\end{array}
\label{interMSSM}
\eeq
and $I_{j,O} = 0$. By the general rules of the previous section, this  translates into the particle content of table \ref{MSSM}, where we have chosen the particular value $g=3$.
\TABLE{\renewcommand{\arraystretch}{1.2}
\begin{tabular}{|c|c|c|c|c|c|}
\hline Sector &
 Matter fields  & $SU(3) \ti SU(2)_L $  &  $Q_a$  & $Q_j$  & $Q_{Y}$ \\
\hline\hline (ab) & $Q_L$ &  $3(3,2)$ & 1 & 0 & 1/6 \\
\hline (aj) & $U_R$   &  $3( {\bar 3},1)$ & -1  &  1  & -2/3 \\
\hline (aj') & $D_R$  &  $3(\bar{3},1)$ &  -1 & -1  & 1/3  \\
\hline (jb) & $L$  &  $3(1,2)$ & 0 & 1  &  -1/2  \\
\hline (jj') & $E_R$   &  $3(1,1)$ &  0  & -2  &  1  \\
\hline 
\end{tabular}
\label{MSSM}
\caption{\small Chiral MSSM spectrum and $U(1)$ charges obtained from the intersection products (\ref{interMSSM}), for the particular choice $g=3$. The hypercharge generator is defined as $Q_{Y} = \frac 16 Q_a - \frac 12 Q_j$.}}

Notice that after the recombination (\ref{recomb1}) we are left with {\em just the MSSM gauge group $SU(3) \times SU(2)_L \times U(1)_Y$}. Indeed, as in the LR parent model, there is one extra $U(1)$ symmetry which becomes massive after taking into account the D-brane couplings to the closed string sector, and we are left with just one Abelian factor, to be identified with the hypercharge. The massive $U(1)$ symmetry, which is nothing but the Baryon number, remains as a perturbative global symmetry of the effective Lagrangian.

The particle content of table \ref{MSSM} is also minimal in its chiral spectrum. Indeed, notice that {\em right-handed neutrinos are not present} in table \ref{MSSM}, and this nicely matches with the fact that by the recombination (\ref{recomb1}) we have broken the $U(1)$ lepton number, which was associated with the D-brane stack $d$, even as a global symmetry of the theory. Indeed, computations analogous to those in \cite{cim2} show that, in field theory language, the recombination (\ref{recomb1}) can be identified with giving a v.e.v. to the scalar fields in  $L_R$.\footnote{Of course, even if we give a v.e.v. to $L_R$ we may still have massless right-handed neutrinos $N_R$ in our theory. These may show up as chiral multiplets in the adjoint of $U(1)_j$. That is, we would associate $N_R$ as modulinos of the brane $j$. However, since these chiral fields do not arise from intersection products any longer, they are not protected by gauge invariance and once supersymmetry is broken by, e.g., fluxes they would generically get a mass.}

One may wonder where does the Higgs sector arise in this MSSM D-brane model. Keeping track of the Higgs fields before and after the $SU(2)_R \times U(1)_{B-L}$ breaking, one concludes that the $H_u$ and $H_d$ chiral multiplets must arise from the $bj$ sector. This may seem puzzling at first sight, since the intersection product $I_{jb} = g$ only seems to have room for $g$ generations of left-handed leptons. There is, however, no contradiction with our results, since the topological number $I_{jb}$ only give us information about the {\em net} chirality of the sector $jb$. As we already mentioned, the Higgs sector as a whole is non-chiral, and in this particular model $H_u$ and $H_d$ arise as a massive vector-like pair in the $jb$ sector. The mass of this vector like pair, which is nothing but the MSSM $\mu$-term, would depend on the original LR $\mu$-term as well as the recombination v.e.v. $\langle L_R \rangle$.\footnote{Fluxes provide extra sources of $\mu$-term \cite{ciu2}, so it would be nice to see how fluxes changes this picture.}

%%%%%%%%%%%%%%%%%%%%%%%%
\EPSFIGURE{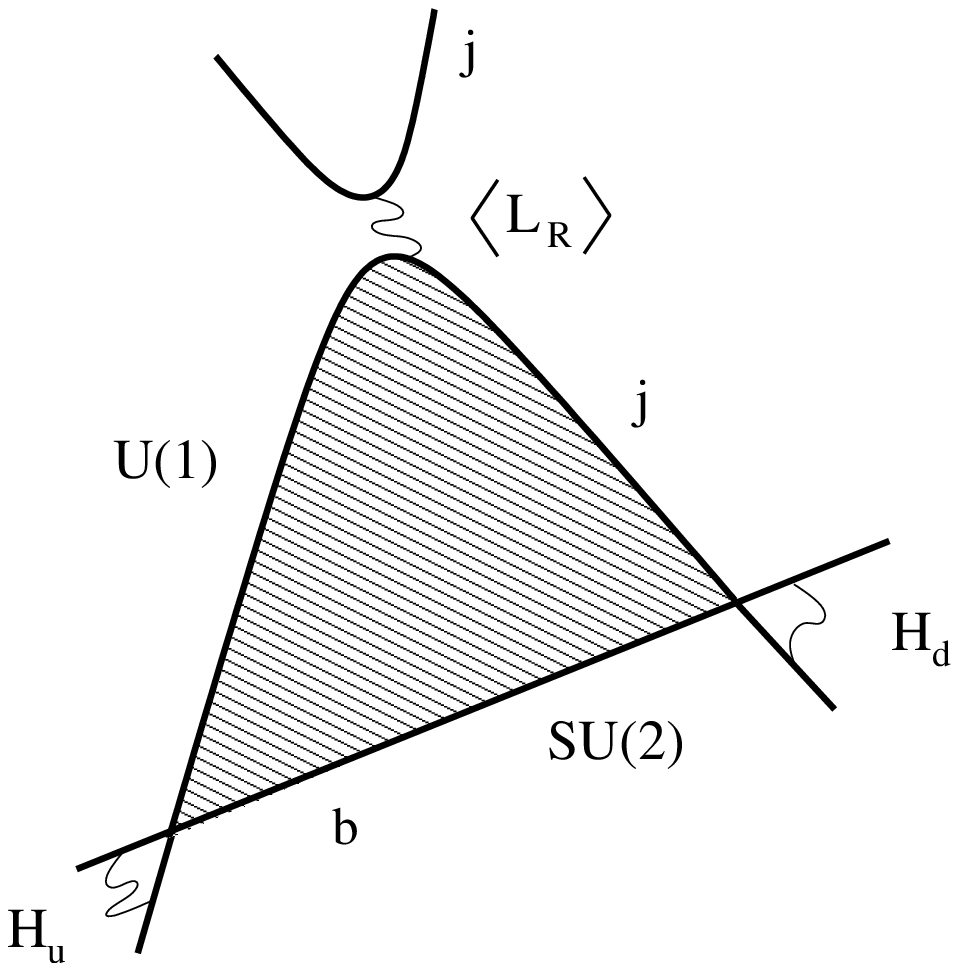, width=3in}
{\label{recomb}
Schematic picture of D-brane recombination (T-dual magnetised D-branes). After giving a v.e.v. to $L_R$, stacks $c$ and 
$d$ have recombined into a third one $j$.
If the recombination is soft enough, the number of 
chiral fermions at the intersections will not vary. 
However, $H_u$ and $H_d$ will get a mass term from worldsheet instantons.}
%%%%%%%%%%%%%%%%%%%%%%%%

These facts are probably easier to visualise in the T-dual picture of intersecting D6-branes. In this picture the stacks $b$ and $j$ are given by two Special Lagrangian 3-cycles, $\Pi_b$ and $\Pi_j$, of $\T^6/(\inte_2 \times \inte_2)$. Although $\Pi_b$ can be described by a simple product of cycles
%\footnote{Strictly speaking, $\T^6/(\inte_2 \times \inte_2)$ has no 1-cycles. By this we mean that $\Pi_b$ comes from a product of three 1-cycles in the covering space $\T^6$, properly identified under the orbifold action.} 
this is not the case for $\Pi_j$, which is a non-factorisable 3-cycle. This implies that the intersection number $I_{jb} = [\Pi_j] \cdot [\Pi_b]$ is not necessarily equal to the number of intersections $\#(\Pi_j \cap \Pi_b)$. Indeed, is easy to see that upon a `soft recombination' (\ref{recomb1}) (i.e., with small $\langle L_R \rangle$), the actual number of intersections will  not change, i.e., $\#(\Pi_j \cap \Pi_b) = \#(\Pi_c \cap \Pi_b) + \#(\Pi_d \cap \Pi_b) = 2 + g$, while the intersection number is only $I_{jb} = g$.\footnote{In order to simplify our discussion, we are not being totally precise with the terminology. Instead of `number of intersections' $\#(\Pi_j \cap \Pi_b)$, we should instead talk about the well-defined mathematical notion of Floer Homology groups $HF^i(\Pi_j,\Pi_b)$. We will not dwelve on these technical aspects in this paper.} All this means that, between $\Pi_j$ and $\Pi_b$, there are $g$ `topological' intersections, giving rise to the chiral multiplets $L^i$, and then two chiral multiplets of opposite chirality and same quantum numbers as $L^i$, 
%\footnote{More precisely, we have these two chiral multiplets and their CPT conjugates.} 
also localised at two extra intersections, which give us the Higgs fields $H_u$ and $H_d$. This vector like pair gets a mass term by worldsheet instantons connecting both intersections \cite{cim2}. This has been schematically illustrated in figure \ref{recomb}.

In this same spirit, is easy to see that the electroweak symmetry breaking of the MSSM down to $SU(3) \times U(1)_{\rm e.m.}$ can also be understood as a D-brane recombination, this time of the form $b+j \raw e$. It is easy to check that, after this second D-brane recombination, we are left with a gauge group $SU(3) \times U(1)_{\rm e.m.}$ and that the intersection products between the stacks $a$ and $e$ vanish. We thus recover no net chiral fermion charged under the EW-broken MSSM gauge group, which means that all MSSM fermions may have finally obtained a mass after both D-brane recombinations.

Notice that this MSSM-block is extremely simple in many aspects, and particularly minimal regarding both gauge groups and chiral matter. However, it still does not provide us with just the particle content of the MSSM. Indeed, we have plenty of non-chiral matter, like the three massless $SU(3)$ adjoints fields. These and other non-chiral fields are related to D-brane positions and Wilson lines, i.e., open string moduli of the model. We then find that the particle content beyond the MSSM that we find in this local construction is actually nothing but the old moduli problem of supersymmetric string vacua. 

As explained in the previous section, we expect RR and NSNS fluxes to fix most of these moduli, even in the case of open strings. We thus turn to the task of embedding this local MSSM-like model in a global flux compactification.

\section{Examples}

In this section we finally construct some $\cn=1$ and $\cn=0$ tadpole-free\footnote{In the following, by tadpole-free we mean string models {\em free of  both RR and NSNS tadpoles}, which is not a common feature of $\cn =0$ compactifications. More precisely, in the $D=10$ supergravity approximation we are satisfying both Bianchi identities and Einstein's equations with a $D=4$ Poincar\'e invariant vacuum.} chiral flux vacua. Given the material presented in previous sections our strategy should be clear by now. We consider Type IIB string theory, compactified on $\T^6/(\inte_2 \times \inte_2)$ with discrete torsion, and try to build a D-brane model including RR and NSNS 3-form fluxes and the D-brane MSSM block of table \ref{MSSMblock}. In order to be a consistent string theory construction, we should satisfy the RR tadpole conditions (\ref{tadpoles}), (\ref{tadpolesK}) and (\ref{tadpoleflux}). If, in addition, we want to have a tadpole-free model then we should also satisfy conditions (\ref{susy}).

As we will presently see, the presence of non-trivial fluxes in these tadpole-free constructions implies that we must consider magnetised D9-branes as well. We will first discuss the model building possibilities that magnetised D9-branes add, and then give some examples which show how the $\cn=1$ and $\cn=0$ flux vacua mentioned above can be achieved.

\subsection{Adding fluxes and extra D-branes}

The authors of \cite{blt,cu} found a clear obstruction
in building $\cn=1$ flux models. The D3-brane charge and tension that a non-trivial 3-form flux would carry already overshoots the O3-plane negative charges\footnote{This is actually only valid for the $\inte_2 \times \inte_2$ orientifold with discrete torsion. An analogue obstruction can be found in the case without discrete torsion. See \cite{blt} and next section.}. Indeed, the quantisation conditions (\ref{quant}) with $N_{\rm min} = 8$ imply that $N_{\rm flux} = n \cdot 64$, $n \in {\bf N}$. The D3-brane RR tadpole condition (\ref{tadpoleflux}) then reads
\beq
\sum_\a N_\a n_\a^1 n_\a^2 n_\a^3 + n \cdot 32 = 16,
\label{obstruct}
\eeq
and one finds that, for $n \neq 0$, $\sum_\a N_\a n_\a^1 n_\a^2 n_\a^3 < 0$. This in particular means that, if one foresees a simple model where we only have D3 and D7-branes with magnetic fluxes, we should eventually add anti-D3-branes in order to give a negative contribution to the r.h.s. of (\ref{obstruct}) and satisfy RR tadpole conditions. These anti-D3-branes would not only break supersymmetry but also generate a large NSNS tadpole, and hence the Einstein's equations with a $D=4$ Poincar\'e invariant metric would never be satisfied.

In any case, it seems clear that in order to satisfy RR tadpoles with a non-trivial flux we must add objects carrying anti-D3-brane charge. While it may seem difficult to do this without breaking supersymmetry, it actually turns out to be possible. Indeed, is easy to see that magnetised D9-branes may carry either anti-D3-brane or anti-D7$_i$-brane charge, while still preserving the $\cn=1$ supersymmetry of the orientifold background. The T-dual version of this striking fact was already pointed out in \cite{csu}, where some explicit models with this property were constructed.

For instance, a D9-brane $\a$ with magnetic numbers
\beq
(-n_\a^1, m_\a^1) \otimes (-n_\a^2, m_\a^2) \otimes (-n_\a^3, m_\a^3) 
\label{ex1}
\eeq
and with $n_\a^i, m_\a^i > 0$, carries anti-D3-brane charge and D7$_i$-brane charge, $i=1,2,3$. The same applies to its image under $\OR$. For some particular choice of the K\"ahler parameters $\ca_i$ both preserve the same supersymmetry as the orientifold background. For this particular case the $\cn=1$ condition reads
\beq
\sum_i {\rm tan}^{-1} \left( {m_a^i \ca_i \over n_a^i} \right) = \pi.
\label{susy2}
\eeq
On the other hand, an anti-D9-brane $\b$ with numbers
\beq
(n_\b^1, m_\b^1) \otimes (n_\b^2, m_\b^2) \otimes (n_\b^3, -m_\b^3) 
\label{ex2}
\eeq
will carry D3, D7$_1$ and D7$_2$-brane charge, while anti-D7$_3$-brane charge. The supersymmetry conditions for this D-brane are given by (\ref{susy}).

Since one of our main concerns in the present paper is to build $\cn=1$ flux vacua, we will be mainly dealing with magnetised D9-branes of the kind (\ref{ex1}), which allows us to satisfy the RR constraint (\ref{obstruct}) without breaking supersymmetry.

\subsection{to achieve $\cn=1$ and $\cn=0$ chiral flux vacua}

Let us now illustrate the above ideas with an explicit example. In table \ref{Ymodel} we present a magnetised D-brane model which satisfies the necessary requirements to accommodate both the MSSM local model of the previous section and non-trivial 3-form fluxes, while still satisfying RR and NSNS tadpole conditions. Indeed, it is easy to check that these magnetic numbers satisfy the conditions (\ref{tadpoles}) and (\ref{tadpolesK}), by simply imposing $g^2 + N_{f} = 14$. Notice that this give us an upper bound for the number of generations, namely $g \leq 3$. 

The spectrum of this model includes the local Left-Right MSSM of the previous section, as well as new gauge and chiral sectors coming from the extra two stacks of magnetised D9-branes and the $8N_f$ D3-branes. More precisely, the gauge group of this model is% given by
\beq
SU(3) \times SU(2) \times SU(2) \times U(1)_{B-L} \times \left[ U(1)'  \times USp(8N_f) \right],
\label{gauge}
\eeq
where $U(1)' = [U(1)_a + U(1)_d] - 2g\, [U(1)_{h_1} - U(1)_{h_2}]$ is the only Abelian factor that, besides $U(1)_{B-L}$, survives the generalised Green-Schwarz mechanism. The $USp(8N_f)$ gauge group will only remain as such when all the D3-branes are placed on top of an orientifold singularity. Eventually, by moving them away it can be Higgsed down to $U(1)^{2N_f}$. Of course, the new D-brane sectors will also imply new chiral matter, some of it charged under the Left-Right MSSM gauge group. We will explain below how to deal with these chiral exotics.

Let us first focus on completing our $\cn=1$ embedding. For this we need to cancel NSNS tadpoles as well, which amounts to require
\beq
\begin{array}{c}\vspace*{.15cm}
\ca_2 = \ca_3 \\
{\rm tan}^{-1} (\ca_1/2) + {\rm tan}^{-1} (\ca_2/3) + {\rm tan}^{-1}(\ca_3/4)=\pi
\end{array}
\label{susyYmodel}
\eeq
simultaneously. These two conditions come from (\ref{susy}) and (\ref{susy2}), respectively, and fix the K\"ahler parameters $\ca_i$ in terms of the overall volume $\ca_1\ca_2\ca_3$. This `fixing' of moduli should not be thought of as a dynamical stabilisation process\footnote{There is some confusion about this fact in the literature. For a proper discussion of this point see \cite{csu}.}, as happens, e.g., with complex structure moduli in presence of fluxes, but simply a condition imposed by hand.

\TABLE{\renewcommand{\arraystretch}{1.2}
\begin{tabular}{|c||c|c|c|}
\hline
 $N_\a$  &  $(n_\a^{1},m_\a^{1})$  &  $(n_\a^{2},m_\a^{2})$
&  $(n_\a^{3},m_\a^{3})$ \\
\hline\hline $N_a = 6$ & $(1,0)$ & $(g,1)$ & $(g,-1)$  \\
\hline $N_b=2$ & $(0,1)$ &  $ (1,0)$  & $(0,-1)$ \\
\hline $N_c=2$ & $(0,1)$ &  $(0,-1)$  & $(1,0)$  \\
\hline $N_d = 2$ & $(1,0)$ & $(g,1)$ & $(g,-1)$  \\
\hline
\hline $N_{h_1}= 2$ & $(-2,1)$  & $(-3,1)$ & $(-4,1)$ \\
\hline $N_{h_2}= 2$ & $(-2,1)$ & $(-4,1)$ & $(-3,1)$ \\
\hline $8N_{f} $ & $(1,0)$ &  $(1,0)$  & $(1,0)$  \\
\hline \end{tabular}
\label{Ymodel}
\caption{\small D-brane magnetic numbers giving rise to an $\cn=1$ embedding of the MSSM block of table \ref{MSSMblock}. This D-brane model may also admit $\cn=1$ or $\cn=0$ fluxes, depending on the values of $g$ and $N_f$.}} 

When considering fluxes, the first RR tadpole condition in (\ref{tadpoles}) gets modified to (\ref{tadpoleflux}). Since flux quantisation conditions imply that $N_{\rm flux} = n \cdot 64$, $n \in {\bf N}$, this translates into
\beq
g^2 + N_f + 4n = 14.
\label{tadpolefin}
\eeq
which has several solutions. Among them we find:
\begin{itemize}

\item $n = 0$,\quad $g = 3$,\quad $N_f = 5$

\item $n = 1$,\quad $g = 3$,\quad $N_f = 1$

\item $n = 2$,\quad $g = 2$,\quad $N_f = 2$

\item $n = 3$,\quad $g = 1$,\quad $N_f = 1$

\end{itemize}

Even if now we include a non-trivial $G_3$ flux, both RR and NSNS tadpoles will be still satisfied if we demand (\ref{susyYmodel}) and $G_3$ to be ISD, i.e., of the form (\ref{ISD}). That is, in some sense we can trade D3-branes by a BPS-like ISD flux $G_3$ without affecting tadpole conditions. However, the amount of supersymmetry may well vary, since a $(0,3)$ component of $G_3$ will softly break $\cn=1$ to $\cn=0$.

Let us discuss the above solutions in some detail. The first of them corresponds to a 3 generation model where any 3-form flux has been turned off. This example is, indeed, the {\em first $\cn=1$ embedding of the MSSM local model constructed in \cite{yukis}}. The last of these solutions is also very interesting. The quantum of flux $N_{\rm flux} = 3 \cdot 64 = 192$ can be achieved by considering the 3-form flux
\beq
G_3\, =\, \frac{8}{\sqrt{3}}\, e^{-\frac{\pi i}{6}}\, ( d{\ov 
z}_1dz_2dz_3 +
dz_1d{\ov z}_2dz_3 + dz_1dz_2d{\ov z}_3 )
\label{susyflux}
\eeq
which is well quantised at the particular value $\tau_1 = \tau_2 = \tau_3 = \tau = e^{2\pi i/3}$ for the untwisted complex structure moduli and the dilaton. These are indeed the values where those fields get fixed after the scalar potential generated by $G_3$ is minimised \cite{kst,cu}. Notice that the flux (\ref{susyflux}) is a combination of $(2,1)$ 3-forms, and hence the closed string background as a whole preserves $\cn=1$ supersymmetry \cite{granapol}. We thus find that, unlike what previous attempts may have suggested \cite{blt,cu}, it is actually possible to find chiral $\cn=1$ string theory vacua involving 3-form fluxes and magnetised D-branes. The above is indeed the {\em first example of $\cn=1$ chiral type IIB flux compactification}, and the first explicit example of a $D=4$ Minkowski vacuum in string theory where chirality, $\cn=1$ supersymmetry and moduli stabilization by means of fluxes happen at the same time.

Actually, the D-brane content of table \ref{Ymodel} not only allows us to find all these interesting examples, but also provides us with a very nice example of a more realistic model where flux-induced supersymmetry breaking occurs. Indeed, the second solution to (\ref{tadpolefin}) involves a MSSM-like spectrum with 3 generations and $N_{\rm flux} = 64$. The latter can be achieved by
\beq
G_3\, =\,  2 \, ( d{\ov z}_1dz_2dz_3 + dz_1d{\ov z}_2dz_3 + dz_1dz_2d{\ov z}_3 + d{\ov z}_1 d{\ov z}_2 d{\ov z}_3).
\label{nonsusyflux}
\eeq
A particular value of the moduli where such flux is well quantised is given by $\tau_1 = \tau_2 = \tau_3 = \tau = i$. Notice that this give us $g_s = 1$ and the string perturbation theory may seem no longer reliable. It turns out, however, that there are also solutions with arbitrarily small $g_s$. Indeed, as shown in Appendix \ref{moduli}, the scalar potential derived from the flux (\ref{nonsusyflux}) has several flat directions. In particular, it vanishes when one imposes the complex structure moduli and the dilaton to be pure imaginary
\beq
\begin{array}{l} \vspace*{.1cm}
\tau_i = i t_i, \quad \quad t_i \in \real \\
\tau = i/g_s,
\end{array}
\label{const1}
\eeq
and to satisfy the constraint
\beq
g_s t_1 t_2 t_3 = 1.
\label{const2}
\eeq
So, in principle, by varying the parameters $t_i$ one can consider the string coupling to be very close to zero. Of course, $\a'$ corrections may modify this picture \cite{becker}, and lift the flat directions left by (\ref{const1}), (\ref{const2}), dynamically fixing $g_s$. The evaluation of such effects is beyond the scope of the present paper.

\subsection{where chiral exotics can be Higgsed away}

As mentioned before, the addition of the D-brane sectors $h_1$, $h_2$ and $f$, which are necessary to embed the MSSM local model into a global $\cn=1$ compactification, add new gauge groups as well as chiral matter to the low energy spectrum described in the previous section. In general, some of this additional chiral matter will be charged under the MSSM gauge group, and hence will introduce chiral exotics in our spectrum. Nevertheless, as usual in $\cn=1$ theories, we can get rid of most of these exotics by taking appropriate scalar flat directions. In the present context, such flat directions can be engineered from the D-brane perspective by the process of D-brane recombination. Recall that such process was already used in the previous section, in order to break the Left-Right Supersymmetric Standard Model down to the MSSM, and then to break electroweak symmetry.
%mimick EW symmetry breaking.

Let us demonstrate these general ideas with an explicit example. In order to simplify our discussion, we will consider the $\cn=1$ three generation model without fluxes, that is, the solution $n=0$, $g=3$, $N_f=5$  above. We will also consider that stacks $a$ and $d$ are on top of each other and hence we have a Pati-Salam gauge group. This will hardly affect the discussion, but will render our expressions more compact. Finally, we will assume that all the  D3-branes are at the origin, and hence our gauge group includes a $USp(40)$ factor.

The total chiral spectrum of this model is displayed in table \ref{Yspectrum}, including the charges of the chiral matter under the only $U(1)$ factor which is massless. This $U(1)$ is given by the combination $U(1)' = \frac 13 U(1)_a + 2\, [U(1)_{h_1} - U(1)_{h_2}]$, and almost all the chiral matter is charged under it.
\TABLE{\renewcommand{\arraystretch}{1.2}
\begin{tabular}{|c|c|c|c|c|c|c|}
\hline Sector &
 Matter  & $SU(4) \ti SU(2) \ti SU(2) \ti [USp(40)]$  & $Q_a$ & $Q_{h_1}$ & $Q_{h_2}$  & $Q'$   \\
\hline\hline (ab) & $F_L$ &  $3(4,2,1)$ & 1 & 0 & 0 & $1/3$   \\
\hline (ac) & $F_R$   &  $3( {\bar 4},1,2)$ & -1 & 0 & 0 & -$1/3$  \\
\hline (bc) & $H$    &  $(1,2,2)$ & 0 & 0 & 0 & 0 \\
\hline
\hline ($ah_1$) &    &  $6(\bar{4},1,1)$ & -1 & 1 & 0 &  $5/3$  \\
\hline ($ah_2'$) &    &  $6({4},1,1)$ & 1 & 0 & 1 &  -$5/3$  \\
\hline ($bh_1$) &    &  $8(1,2,1)$ & 0 & 1 & 0 & 2  \\
\hline ($bh_2$) &    &  $6(1,2,1)$ & 0 & 0 & 1 & -2  \\
\hline ($ch_1$) &    &  $6(1,1,2)$ & 0 & 1 & 0 & 2  \\
\hline ($ch_2$) &    &  $8(1,1,2)$ & 0 & 0 & 1 & -2  \\
\hline 
\hline ($h_1h_1'$) & & $46(1,1,1)$ & 0 & 2 & 0 & 4 \\
\hline ($h_2h_2'$) & & $46(1,1,1)$ & 0 & 0 & 2 & -4 \\
\hline ($h_1h_2'$) & & $196(1,1,1)$ & 0 & 1 & 1 & 0 \\
\hline ($fh_1$) & & $(1,1,1) \ti [40]$ & 0 & -1 & 0 & 2 \\
\hline ($fh_2$) & & $(1,1,1) \ti [40]$ & 0 & 0 & -1 & -2 \\
\hline
\end{tabular}
\label{Yspectrum}
\caption{\small Chiral spectrum of the three generation Pati-Salam $\cn=1$ chiral model of table \ref{Ymodel}. The Abelian generator of the unique massless $U(1)$ is given by $Q' = \frac 13 Q_a + 2 (Q_{h_1} - Q_{h_2})$.}}
The two exceptions are the Higgs multiplet and the 196 singlets in the $h_1h_2'$ sector of the theory. The latter are of particular interest, since they parametrise a subspace of flat directions in the $\cn=1$ effective theory. Indeed, we can give a non-vanishing v.e.v. to a particular combination of the scalar fields in the 196 chiral multiplets without breaking supersymmetry. In terms of D-brane physics, this is nothing but the D9-brane recombination
\beq
h_1 + h_2' \raw h.
\label{recomb3}
\eeq
More precisely, it amounts to deforming the gauge bundle on the D9-branes, from a direct sum of the Abelian bundles $h_1$ and $h_2'$ to a non-Abelian bundle given by $h$. As usual, the magnetic charges of the new bundle will be given by $[{\bf Q}_h] = [{\bf Q}_{h_1}] +  \OR [{\bf Q}_{h_2}]$.

Notice that giving a v.e.v. to these singlets only breaks  the $U(1)_{h_1} + U(1)_{h_2}$ gauge factor, which was already massive at low energies. Thus, this Higgsing does not affect the low energy gauge group, and in particular the Pati-Salam sector. It does, however, have an important effect on the chiral spectrum of the theory. Indeed, we can compute the chiral spectrum after (\ref{recomb3}) with the charge vector $[{\bf Q}_h]$ and the topological formulae of table \ref{matter}, finding that the final theory has the extremely simple chiral content of Table \ref{Yspectrum2}. As expected from field theory arguments, we still have some chiral exotics given by two doublets of $SU(2)_L$ and two of $SU(2)_R$. Although in principle they are unwanted particles, they will become massive after electroweak symmetry breaking, so a very high mass eigenvalue could in principle make them unobservable at low energies. The phenomenological viability of this spectrum will then crucially depend on the precise Yukawa couplings of this model. We leave these and other phenomenological issues for future work.

\TABLE{\renewcommand{\arraystretch}{1.25}
\begin{tabular}{|c|c|c|c|c|c|}
\hline Sector &
 Matter  & $SU(4) \ti SU(2) \ti SU(2) \ti [USp(40)]$  & $Q_a$ & $Q_{h}$ & $Q'$   \\
\hline\hline (ab) & $F_L$ &  $3(4,2,1)$ & $1$ & 0 & $1/3$   \\
\hline (ac) & $F_R$   &  $3( {\bar 4},1,2)$ & -1 & 0 & -$1/3$  \\
\hline (bc) & $H$    &  $(1,2,2)$ & 0 & 0 & 0 \\
\hline
\hline (bh) &    &  $2(1,2,1)$ & 0 & 1 & 2  \\
\hline (ch) &    &  $2(1,1,2)$ & 0 & -1 & -2  \\
\hline 
\end{tabular}
\label{Yspectrum2}
\caption{\small $\cn=1$ spectrum derived from the D-brane content of table \ref{Ymodel} after D-brane recombination. There is no chiral matter arising from $ah$, $ah'$, $hh'$ or charged under $USp(40)$. The generator of $U(1)'$ is now given by $Q' = \frac 13 Q_a + 2 Q_{h}$.}}

As an important point, notice that the Pati-Salam sector of the theory (i.e., the upper part of the table \ref{Ymodel}) is not involved in the D-brane recombination process, and hence remains as a simple sector of three sets of D7$_1$, D7$_2$ and D7$_3$-branes. In particular, this means that the phenomenological analysis performed for the MSSM local model of the previous section does also apply to the present global embedding. Finally, notice that all the arguments involving D-brane recombination/Higgsing have been performed in the absence of 3-form fluxes. Since these fluxes generically lift some of the flat directions of an $\cn=1$ theory, it would be interesting to see how they affect this picture. We hope to report on these issues in the future.

\section{Extension to other $\inte_2 \times \inte_2$ orientifolds \label{extra}}

So far, in our quest of building chiral flux vacua, we have only considered a particular orientifold background, namely a $\inte_2 \times \inte_2$ orientifold with discrete torsion. Our techniques, however, do also apply to other closed string backgrounds, and the purpose of this section is to illustrate this fact. For simplicity, we will consider another $\inte_2 \times \inte_2$ orientifold of $\T^6$, but this time {\em without} discrete torsion. As was analysed in \cite{blt}, this alternative choice of discrete torsion changes some of the orientifold charges and the flux quantisation conditions, so the model building rules differ from the $\inte_2 \times \inte_2$ considered previously. Nevertheless, we again manage to find $\cn=1$ flux vacua in this background.

Actually, in the absence of fluxes, this background is T-dual to the $\inte_2 \times \inte_2$ type I models discussed in \cite{aadds}. In this reference several models, based on $\inte_2 \times \inte_2$ and other backgrounds, were constructed. The purpose of these examples was to illustrate the phenomenon of `brane supersymmetry breaking' \cite{bssb}. Roughly speaking, these are constructions where the closed string sector of the theory is $\cn=1$ supersymmetric but, due to the RR tadpole consistency conditions, there is no naive $\cn=1$ consistent open string sector one can build. For instance, in the particular case of the $\inte_2 \times \inte_2$ orientifold without discrete torsion one cannot build a supersymmetric model by considering D-branes without magnetic fluxes, as one can do in the case with discrete torsion \cite{bl}. However, as we presently show, there exist $\cn=1$ solutions which involve magnetised D-branes and, moreover, they also admit non-trivial fluxes.

Let us first review the model building rules for magnetised D-branes and 3-form fluxes in a $\T^6/(\inte_2 \times \inte_2)$ orientifold without discrete torsion, following \cite{blt}. Most part of the discussion parallels the rules of the $\inte_2 \times \inte_2$ orientifold with discrete torsion, so let us just point out the differences. First, in this alternative background there is at least one class of orientifold planes with positive charge and tension. For simplicity, we will choose the 64 O3-planes to be exotic O3$^{(+,+)}$'s with positive charge and tension, while the O7$_i$-planes, $i=1,2,3$ will have the usual negative charges. This will change the magnetic numbers of the orientifold plane content of the theory, which instead of (\ref{oriclass}) will read
\beqa \nonumber
 [{\bf Q}_O] & = & - [(1,0)\ (1,0)\ (1,0)] \ +\  [(1,0)\ (0,1)\ (0,-1)] \\
& + & [(0,1)\ (1,0)\ (0,-1)] \ +\  [(0,1)\ (0,-1)\ (1,0)].
\label{oriclass2}
\eeqa

Second, following the strategy in \cite{blt} we will consider `bulk' branes, which come in four copies, and not the fractional branes which have been considered so far. This implies that from a stack of $N_a$ D-branes not fixed by the orientifold action we get a $U(N_a/4)$ gauge group, and that several factors of 4 appear in the chiral spectrum. This massless spectrum has been summarised in table \ref{matter3}.
\TABLE{\renewcommand{\arraystretch}{1.25}
\begin{tabular}{|c|c|}
\hline
\hspace{1cm} {\bf Sector} \hspace{1cm} &
\hspace{1cm} {\bf Representation} \hspace{1cm} \\
\hline\hline
$aa$   &  $U(N_a/4)$ vector multiplet  \\
       & 3 Adj. chiral multiplets   \\
\hline\hline
$ab+ba$   & $4 I_{ab}$ $(\fund_a,\antifund_b)$ chiral multiplets  \\
\hline\hline
$ab'+b'a$ & $4 I_{ab'}$ $(\fund_a,\fund_b)$ chiral multiplets  \\
\hline\hline
$aa'+a'a$ & $2 (I_{aa'} - 2 I_{a,O}) \;\;
\Ysymm\;\;$ chiral  multiplets \\
          & $2 (I_{aa'} +  2 I_{a,O}) \;\;
\Yasymm\;\;$ chiral multiplets \\
\hline
\end{tabular}
\label{matter3}
\caption{\small Massless spectrum for $\cn=1$ magnetised D-branes in the $\IT^6/(\inte_2\times \inte_2)$ $\OR$ orientifold without discrete torsion. $I_{a,O}$ is now computed from the orientifold plane charge class (\ref{oriclass2}).}}

On the other hand, a stack of $N$ D-branes on top of an O-plane will no longer give us a $USp(N)$ gauge group. Instead, $N_a$ D3-branes on top of a O3$^{(+,+)}$ will give us a $USp(N_a/2)^4$ gauge group, whereas $N_a$ bulk D7$_i$-branes without magnetic fluxes will yield a $U(N_a/2) \times U(N_a/2)$ gauge group. We present such spectra in tables \ref{matter4} and \ref{matter5}.
\TABLE{\renewcommand{\arraystretch}{1.25}
\begin{tabular}{|c|c|}
\hline
\hspace{1cm} {\bf Sector} \hspace{1cm} &
\hspace{1cm} {\bf Representation} \hspace{1cm} \\
\hline\hline
$aa$   &  $USp(N_a/2)^4$ vector multiplets  \\
       &  $(\underline{\fund\,,\fund\,,1\,,1})$ chiral multiplets   \\
\hline\hline
$ab+ba$   & $I_{ab}$ $(\underline{\fund_a,1\,,1\,,1\,};\antifund_b)$ chiral multiplets  \\
\hline
\end{tabular}
\label{matter4}
\caption{\small Massless spectrum for D3-branes invariant under the orientifold action. The underline stands for all possible permutations of the representations.}}

\TABLE{\renewcommand{\arraystretch}{1.25}
\begin{tabular}{|c|c|}
\hline
\hspace{1cm} {\bf Sector} \hspace{1cm} &
\hspace{1cm} {\bf Representation} \hspace{1cm} \\
\hline\hline
$aa$   &  $U(N_a/2) \times U(N_a/2)$ vector multiplets  \\
       &  $(\fund\,,\fund)$ $+$ $(\fund\,,\antifund)$ $+$  
	  $(\antifund\,,\fund)$ $+$ $(\antifund\,,\antifund)$ \\
       &  $\left(\Yasymm\,,1\right)$ $+$ $\left(\bYasymm\,,1\right)$ $+$ $\left(1\,,\Yasymm\right)$ $+$ $\left(1\,,\bYasymm\right)$ \\ & chiral multiplets   \\
\hline\hline
$ab+ba$   & $I_{ab}$ $\left[(\underline{\fund_a\,,1}\,;\antifund_b)\ + \ (\underline{\antifund_a\,,1}\,;\antifund_b)\right]$ chiral multiplets  \\
\hline
\end{tabular}
\label{matter5}
\caption{\small Massless spectrum for D7$_i$-branes, $i=1,2,3$, invariant under the orientifold action.}}

Third, since the collapsed cycles of the $\inte_2 \times \inte_2$ orbifold do not contain any additional 3-cycles, the integral homology basis of such orbifold, and hence the quantisation conditions for the 3-form fluxes, change with respect to the case with discrete torsion \cite{blt}. As a result, in (\ref{quant}) we must consider $N_{\rm min} = 4$  and hence we can write $N_{\rm flux} = n \cdot 16$, $n \in {\bf N}$ for ISD fluxes. All this changes the RR tadpole conditions, which in the the present context read
\beq
\sum_\a N_\a n_\a^1 n_\a^2 n_\a^3 + \oh N_{\rm flux} = - 16,
\label{tadpoleflux2}
\eeq
as well as the three lower conditions in (\ref{tadpoles}). In addition, we may consider extra K-theory constraints, which may be read, e.g., from the global $USp(2)$ anomalies of this chiral spectrum. In the present context these are
\beq
\sum_\a N_\a m_\a^1 m_\a^2 m_\a^3\  \in\  8 \inte.
\label{tadpolesK2}
\eeq
There may be, however, additional K-theory constraints which, since we are considering only bulk D-branes, are invisible to the low energy spectrum presented above. Finally, the supersymmetry conditions are identical to the case with discrete torsion.

From condition (\ref{tadpoleflux2}) it is easy to see that satisfying RR tadpoles implies that some D-branes must carry anti-D3-brane charge.\footnote{Recall that we are considering ISD fluxes, and hence $N_{\rm flux} \geq 0$.} This is true even in the absence of fluxes, and clearly comes from the fact that we have an O3$^{(+,+)}$-plane with positive charge and tension. As a result, in a D-brane model with only (anti)D3-branes and D7-branes, supersymmetry breaking is enforced by RR tadpole conditions.

Notice that the same kind of situation was found in the previous section, when we wanted to include 3-form fluxes in the $\inte_2 \times \inte_2$ orientifold with discrete torsion. The same kind of strategy for finding $\cn=1$ models applies here. Namely, we can consider D9-branes of the form (\ref{ex1}) which carry anti-D3-brane charge and still preserve $\cn=1$ supersymmetry. Indeed, a simple D-brane model showing this fact is given by the magnetic numbers
\beq
16\ (-1,1) \otimes (-1,1) \otimes (-1,1)
\label{simplest}
\eeq
and it is easy to see that it defines an $\cn=1$ vacuum of the theory, provided the choice of K\"ahler parameters such that $\sum_i tan^{-1} (\ca_i) = \pi$. This simple D-brane content yields an $SU(4)$ gauge group with three adjoint multiplets and 32 chiral multiplets in the ${6}$ representation. This is probably the simplest $\cn=1$ vacuum of the present $\inte_2 \times \inte_2$ orientifold, but it is not very appealing from the phenomenological point of view, and moreover it does not admit the presence of fluxes.

\TABLE{\renewcommand{\arraystretch}{1.2}
\begin{tabular}{|c||c|c|c|}
\hline
 $N_\a$  &  $(n_\a^{1},m_\a^{1})$  &  $(n_\a^{2},m_\a^{2})$
&  $(n_\a^{3},m_\a^{3})$ \\
\hline\hline 
\hline $N_{h}= 8$ & $(-2,1)$  & $(-2,1)$ & $(-2,1)$ \\
\hline $8N_{f} $ & $(1,0)$ &  $(1,0)$  & $(1,0)$  \\
\hline \end{tabular}
\label{newmodel}
\caption{\small D-brane magnetic numbers giving rise to $\cn=1$ flux vacua.}}
An $\cn=1$ model allowing fluxes is given by the D-brane content shown in table \ref{newmodel}. The RR tadpole conditions are automatically satisfied except for (\ref{tadpoleflux2}), which is given by
\beq
 N_f + n = 6.
\label{tadpolefin2}
\eeq
and hence admits several $\cn=1$ flux vacua. 
\TABLE{\renewcommand{\arraystretch}{1.25}
\begin{tabular}{|c|c|}
\hline Sector & $SU(2) \ti USp(4N_f)^4$ \\
\hline(hh) & $3(3;1,1,1,1)$ \\
\hline (hh') & $76(3;1,1,1,1) + 180(1;1,1,1,1)$, \\
\hline (hf) & $(2,\underline{4N_f,1,1,1})$ \\
\hline 
\end{tabular}
\label{newspectrum}
\caption{\small $\cn=1$ spectrum derived from the D-brane content of table \ref{newmodel}.}}
The gauge group of this model is given by $SU(2) \times USp(4N_f)^4$, the $U(1)$ factor from the initial $U(2)$ being actually massive. The $\cn=1$ spectrum is presented in table \ref{newspectrum}. This model is clearly not realistic and, strictly speaking, is not even chiral. This probably comes from the fact that we are only considering bulk D-branes in our construction. A more promising approach seems to consider fractional D-branes, which moreover have the potential to yield chiral representations in multiplicities different than four. In any case, these examples show how our model building techniques allow to construct $\cn=1$ flux vacua in other closed string backgrounds. Hopefully, this will open up new avenues in string theory model building.

\section{Conclusions and Outlook}

In this paper we have developed model building techniques for constructing $\cn=1$, $D=4$ chiral vacua of flux compactifications in Type IIB string theory. More precisely, we have focused on the magnetised D-brane setup of \cite{blt,cu} and we have constructed $\cn=1$ flux models for both choices of $\T^6/({\bf Z}_2 \times {\bf Z}_2)$ orientifolds: with or without discrete torsion. We have discussed the consistency conditions that such models must satisfy and, in particular, derived some additional K-theory constraints on them. These general ideas have been illustrated with the explicit construction of $\cn=1$ and $\cn=0$ global models, which admit both chirality and Poincar\'e invariance. Unlike many other examples in the literature, the $\cn=0$ models are free of NSNS tadpoles (to first order) and its associated instabilities. 

The first example of these type of models was recently presented in \cite{MS}, involving not only fluxes, but also a low energy spectrum that contains the MSSM with three generations of chiral matter. Part of the purpose of this paper is to elaborate on the construction on such model, as well as to describe its phenomenological features. The construction is mainly inspired by the intersecting D-brane world scenario and, as such, shares many of the interesting features of intersecting D-brane models. In particular, we show how the processes of symmetry breaking from a Left-Right model to the MSSM, as well as electroweak symmetry breaking, can be nicely realised in this context in terms of D-brane recombination. 

In addition, these explicit models provide a good starting point for studying the phenomenological consequences of $D=4$ chiral flux compactifications. For instance, the general pattern of SUSY-breaking soft terms deduced from \cite{lrs,ciu2} can be applied to these compact examples. A preliminary analysis naively suggest a typical scale of order $\a'/\sqrt{Vol(T^6)}$ for these terms, which favours an intermediate string scale $M_s = 10^{11} GeV$. However, a non-trivial, inhomogeneous warp factor may change this situation. It would be interesting to see if this is indeed the case. The pattern of soft terms may also depend on the precise way in which the remaining moduli are stabilised.\footnote{We thank Kiwoon Choi and Hans-Peter Nilles for emphasising to us this point.} We also note that there are two sources of the $\mu$ term in these models: the one described in Section 3 and the contribution from the fluxes \cite{ciu2}. It is worth exploring if this fact may have interesting implications for electroweak symmetry breaking. 

Furthermore, these models contain the basic ingredients of the recent proposal for constructing de Sitter vacua from string theory \cite{kklt}, so it would be very nice to find examples of this scenario which could simultaneously lead to de Sitter geometries and phenomenologically interesting vacua. Let us point out, however, that since our constructions are based on toroidal orbifolds, it is not obvious how, e.g., warped throats can be realised in our models. These difficulties could be easily overcomed in Calabi-Yau compactifications including magnetised D-branes and fluxes which, in any case, is a natural generalisation of our constructions.

Actually our construction is, a priori, an example of the general class of flux compactifications studied in \cite{gkp}. Notice, however, that the key ingredient that allows us to construct $\cn=1$ chiral flux vacua is the presence of pairs of $D9-\overline{D9}$-branes in our models, which in principle was not considered in \cite{gkp}. This $D9-\overline{D9}$ system would usually form a non-BPS SUSY breaking brane but, due to the presence of magnetic fields in their worldvolume, do preserve the same supersymmetries as the orientifold background. In addition, such object carries charges of D7-brane and {\em anti}-D3-brane. It would be interesting to understand the properties of such $D9-\overline{D9}$ system in a general flux compactification. Moreover, these objects carry a discrete $\inte_2$ K-theory charge, as we have derived by using the brane probe arguments in \cite{angel}. Notice that this is a field theory argument, so it would be interesting to derive these results directly from first principles.

The inclusion of magnetised $D9-\overline{D9}$-branes also allows us to construct a new class of $\cn=1$ D-brane models, which were actually believed not to exist. These are some compactifications with the property of `brane supersymmetry breaking'. More precisely, in some orientifold examples one finds supersymmetric closed string vacua where the RR tadpole conditions do not allow any naive $\cn=1$ D-brane content. By considering magnetised $D9-\overline{D9}$-branes, we show that such $\cn=1$ vacua do actually exist, at least in a particular $\T^6/(\inte_2 \times \inte_2)$ background T-dual to some models in \cite{aadds}. It would be interesting to see if similar techniques can be applied to obtain new $\cn=1$ D-brane models in other backgrounds of this kind.

On a more theoretical side, it would be interesting to investigate the F-theory lift of these $\cn=1$ and $\cn=0$ compact models, both being chiral examples of the warped metric solutions found in \cite{gkp}. Since our constructions are particularly simple, one can use them to explore the non-K\"ahler geometries involved in their heterotic \cite{drs} and type IIA \cite{kahler} duals.

Finally, there has also been much discussions recently about the possibility of a string landscape, especially in the framework of flux compactification. Our results provide a concrete proof of concept that vacua with features of the MSSM can exist in this context. However, we note that a lot of the flux vacua (which are included in analysing the statistics) do not have the realistic properties of the Standard Model. Therefore, the conclusion one would draw from such statistical analysis may be different when one convolves the statistics with the criteria of a realistic particle physics spectrum.

To sum up, we have constructed some examples of string vacua that gather several essential ingredients for building a fully realistic string theory model, by combining recent insights in flux compactifications and chiral D-brane constructions.  The D-brane sector introduces a non-Abelian gauge group and chiral matter charged under it, allowing for MSSM-like spectra. The 3-form flux background generates a potential for the dilaton and complex structure moduli of the compactification, freezing them to some particular value. Furthermore, the 3-form flux can also induce soft supersymmetry terms for the gauge and chiral sectors of the theory. We find it quite remarkable that all these interesting features can be realised in the same string theory construction. Hopefully, these recent developments have brought us one step closer to understanding the phenomenology consequences of string theory.

\bigskip

\bigskip

\bigskip

\bigskip

\bigskip

\centerline{\bf Acknowledgments}

\bigskip

We wish to thank Bobby Acharya, Kiwoon Choi, Frederik Denef, Hans-Peter Nilles,
Fernando Quevedo and specially Luis Ib\'a\~nez and Angel Uranga for useful comments and discussions. This work was supported in part by NSF CAREER Award No.~PHY-0348093, and a Research Innovation Award from Research Corporation. We also thank Aspen Center for Physics for hospitality while 
this work was written.

\newpage

\appendix

\section{Moduli space of $\cn=0$ flux models \label{moduli}}

In this appendix we derive the minima of the scalar potential $V_{eff}$ induced by the 3-form flux (\ref{nonsusyflux}). As we will see, this potential does not fix the complex structure moduli and complex dilaton completely, but has some flat directions which allow to choose an arbitrarily small string coupling constant. We will basically follow the approach used in \cite{kst}, which we will generalise to the case of SUSY-breaking fluxes.

The $G_3$ flux (\ref{nonsusyflux}) can be expressed in terms of their real components $F_3$ and $H_3$ as
\beq
\begin{array}{c}\vspace*{.2cm}
H_3 = - 8\, dy_1 \wedge  dy_2 \wedge dy_3, \quad \quad
F_3 = 8\, dx_1 \wedge  dx_2 \wedge dx_3,
\end{array}
\label{realfluxesap}
\eeq
which clearly satisfy\footnote{In the conventions $\int dx_1 \wedge dx_2 \wedge dx_3 \wedge dy_1 \wedge dy_2 \wedge dy_3 = 1$.} $N_{\rm flux} = 64$ and $G_3 = F_3 - \tau_4 H_3$ for $\tau_j = i$, $j =1,2,3,4$. Here, in order to simplify the notation we are identifying the complex dilaton ($\tau$ in the main text) with the complex modulus $\tau_4$. These 3-form fluxes create a scalar potential which can be derived form the superpotential 
\beq
W = \int \left(F_3 - \tau_4 H_3 \right) \wedge \Om = 8 (\tau_1\tau_2\tau_3 - \tau_4)
\label{superap}
\eeq

In fact, as shown in \cite{gkp}, the scalar potential vanishes if $D_j W = 0$, where $j$ runs over the complex structure moduli and the complex dilaton, and $D_j$ is the supergravity covariant derivative. In the case of supersymmetric fluxes, where $W=0$ at the minima of $V_{eff}$, these conditions reduce to solving
\beq
\p_j W = 0
\label{minisusyap}
\eeq
where in the non-supersymmetric case $W \neq 0$ they are equivalent to
\beq
\p_j \left[ e^K |W|^2 \right] = 0
\label{mininonsusyap}
\eeq
where $K$ is the part of the scalar potential that depends on the fields $j$. In our case these fields are $\tau_1, \tau_2, \tau_3, \tau_4$ and 
\beq
K= - ln \left(8 \prod_{j=1}^4 \pim \tau_j \right).
\label{kahler}
\eeq

Now, is easy to see that conditions (\ref{minisusyap}) have no solution for the superpotential (\ref{superap}). On the other hand (\ref{mininonsusyap}), when applied to (\ref{superap}) and (\ref{kahler}), are equivalent to
\beq
\begin{array}{lcr}\vspace*{.2cm}
\bar{\tau}_1 \tau_2 \tau_3 & = & \tau_4 \\ \vspace*{.2cm}
\tau_1 \bar{\tau}_2 \tau_3 & = & \tau_4 \\ \vspace*{.2cm}
\tau_1 \tau_2 \bar{\tau}_3 & = & \tau_4 \\
\bar{\tau}_1 \bar{\tau}_2 \bar{\tau}_3 & = & \tau_4 \\
\end{array}
\label{condap}
\eeq
which in turn imply $\tau_i \tau_j \in \real$. It is easy to see that, if we impose $\tau_j$ to be pure imaginary, we get conditions (\ref{const2}).

\end{document}